\documentstyle[fleqn,aaspptwo,tighten,11pt]{article}
\newcommand{\disp}[2]{\mbox{$\sigma#2_{#1}$}}
\newcommand{\dispsq}[2]{\mbox{$\sigma#2_{#1}^2$}}

\newcommand{\etal}{\mbox{{\it et al. }}}
\newcommand{\Msun}{\mbox{$M_{\odot}\;$}}

\newcommand{\MSpccub}{\mbox{$M_{\odot}/{\rm pc}^3\;$}}
\newcommand{\MoverL}{\mbox{${\cal M/L}\;$}}
\newcommand{\as}{\mbox{$''$}}

\newcommand{\Kz}[1]{\mbox{$K_z{#1}$}}

\newcommand{\ddRoR}   {\mbox{$\frac{1}{R }\;\frac{\partial  }{\partial  R}$}}

\newcommand{\ddRoT}{\mbox{$\frac{1}{R}\;\frac{\partial}{\partial \theta}$}}

\newcommand{\WIDTH}{$W_{gas}$($\approx FWHM/2.35$, as defined by eqn. 
   [\protect\ref{eq:gas.width.approximate}])}
\newcommand{\GAMMA}{(defined by eqn. [\protect\ref{eq:gamma.beta.definition}])}
\newcommand{\mc}[1]{\multicolumn{4}{c|}{\mbox{$#1$}}}
\newcommand{\EqnNum}[1]{\newcounter{#1} \setcounter{#1}{\value{equation}} }

\begin{document}

\title{On the usage of Flaring Gas Layers to determine the \\
Shape of Dark Matter Halos}

\begin{center}
To appear in the August 1995 issue of {\em The Astronomical Journal}
\end{center}

\author{Rob P. Olling}
\affil{Columbia University}
\authoraddr{538 West 120$^{th} street box 42, New York, NY10027}

\newcounter{tabcnt}
\newcounter{appAfor}
\newcounter{temp}
\newcounter{actual.number}
\setcounter{appAfor}{0}
\setcounter{temp}{0}
\setcounter{figure}{0}
\setcounter{tabcnt}{0}

\small


\begin{abstract}

I present a new method of deriving the shape of the dark matter (DM)
halos of spiral galaxies.  The method relies on the comparison of model
predictions with high spectral and spatial resolution HI observations of
the gas layer.  So far, determinations of the flaring of the gas layer
(i.e.  the increase of the thickness with galactocentric radius) have
been used to determine the mass-to-light ratio, \MoverL, of the stellar
disk of several edge-on galaxies.  In this paper I describe a method
which can be used to determine the shape of DM-halos.  This technique
will be applied in a forthcoming paper. 

I show that the model predictions of the gas layer width are best
calculated using a global approach, in which the potential arising from
the {\em total} mass distribution of the galaxy is used in the
calculation of the vertical distribution of the gas.  I developed a new
algorithm to calculate the force field of an arbitrary, azimuthally
symmetric, density distribution.  This algorithm is used to calculate the
forces due to the radially truncated stellar disk as well as of the
flaring gas layer. 

I use a simple two-parameter family of disk-halo models which have
essentially the same observed equatorial rotation curve but different
vertical forces.  This mass model is composed of a stellar disk with
constant \MoverL, and a DM-halo with a given axial ratio (Sackett \&
Sparke [ApJ, 1990, 361, 408]).  I approximate the radial force due to
the gaseous disk, and iteratively determine the vertical force due to
the global distribution of the gas. 

In agreement with Maloney [ApJ, 414, 41, 1993] I find that beyond the
Holmberg radius, the thickness of the gaseous disk is sensitive to both
the flattening of the DM-halo and the self-gravity of the gas.  I also
show that the inferred DM-halo flattening is not sensitive to the
particular choice of disk-halo decomposition. 


I show that the determination of the thickness of the gas layer is not
restricted to edge-on galaxies, but can be measured for moderately
inclined systems as well.  Thus, in combination with detailed modeling,
high resolution HI imaging of nearby galaxies with extended HI envelopes
will enable us to determine the shape of the DM halo of these galaxies. 

\end{abstract}

\section{Introduction}

Until recently, little was known about the shape of dark halos :
measurements of the equatorial rotation curve provide a one dimensional
probe to the potential only.  It is possible to construct both spherical
and flat mass models which generate the same rotation curve. 
Therefore, the shape of the dark halo can not be inferred from galactic
rotation curves alone. 

Existing methods to estimate the shape of DM-halos have given mixed
results.  It has been suggested (Dekel \& Shlosman, 1983; Toomre, 1983;
Sparke \& Casertano, 1988) that galactic warps, frequently observed in
the outer parts of gaseous disks, result if the galactic disk is tilted
with respect to the plane of a flattened dark halo.  Hoffner \& Sparke
(1994) investigated the time evolution of such warps and concluded that
only one out of the five systems they studied requires a halo as
flattened as E6.  Polar ring galaxies, early type spirals with material
rings in a plane perpendicular to the galaxy plane, probe the dark halo
potential in two perpendicular directions.  Careful analysis of the
stellar and gas dynamics of the polar ring system NGC 4650A, led Sackett
\& Sparke (1990, hereafter SS90) and Sackett \etal (1994, SRJF94
hereafter) to conclude that the dark halo in this system could be as
flattened as E6-E7 ($q$=0.4 to 0.3).  Recently, Sackett \etal (1994b)
reported the discovery of a rather flattened ($q \approx 0.5$) {\em
luminous} halo around the edge-on spiral NGC 5907, which can account for
the observed rotation curve provided that this luminous material has a
large mass-to-light ratio ($\approx$ 450). 

Like the polar rotation curve, the thickness of the gas layer depends on
the vertical force, \Kz{}, and hence on the shape of the DM-halo
(Maloney 1992; Olling \& van Gorkom 1992; Kundi\'{c} \etal 1993, KHG93;
Maloney 1993, M93).  That the shape of the dark halo influences the
vertical distribution of the gas can be easily understood.  Consider a
round halo, with a certain density distribution, then decrease the
vertical scale-height while maintaining the total mass.  Consequently,
the densities as well as the exerted gravitational forces will increase,
resulting in a thinner HI disk and higher rotation speeds.  In order to
keep the same equatorial rotation curve, one has to deform the DM-halo
in a very specific way (Appendix A, Figure \ref{fig:Rc.rho0.versus.q})
as a result of which the DM-halo densities (at large distances) will be
roughly inversely proportional to the flattening $q \; (=c/a)$.  In
accordance with Maloney (1993), I find that (at large distances) the
thickness of the gas layer is roughly proportional to the square root of
the halo flattening, $q$.  Several authors have employed measurements of
the gas layer thickness to infer the total mass density in the plane of
spiral galaxies.  Van der Kruit (1981) showed that the width of a
galaxy's gaseous disk should increase radially in an exponential
fashion.  From the thickness measurements of the gaseous disk of the
edge-on galaxy NGC 891 by Sancisi \& Allen (1979) van der Kruit
concluded that the \MoverL ratio of the stellar disk does not change
significantly with radius and that the dark halo can not be as flat as
the stellar disk.  Rupen (1991), using more sensitive and much higher
resolution data, found that the width of the gaseous disks of NGC 891
and NGC 4565 increase exponentially, but for NGC 891 with large scatter. 
Our galaxy has been studied in much greater detail by Knapp (1987) and
Merrifield (1992) in HI, and by Malhotra (1994) in CO.  These authors
find that the radial scale-length of the midplane mass density compares
well with the radial scale-length of the stellar disk, implying a
constancy of the mass-to-light ratio for the stellar disk. 

Van der Kruit (1988) concludes that in the outer parts of stellar disks
the local mass density is no longer dominated by the stellar disk so
that the self-gravity of the gas will become important.  Here I extend
van der Kruit's pioneering work by incorporating the stellar and gaseous
disks as well as the DM-halo in a self consistent way : rather than
making a {\em local} approximation to the vertical force, \Kz{} is
calculated from the {\em global} mass distribution of the galaxy.  In \S
\ref{sec-Gaseous-self-gravity} I show that the self-gravity of the gas
can play an important role and must be included in a self-consistent
manner.  Furthermore I will concentrate on the flaring behaviour in the
region beyond the optical disk where the vertical force is dominated by
the DM-halo, and is hence sensitive to its {\em shape}.  In combination
with the newly achievable large sensitivities and the high resolution of
HI synthesis observations (Rupen, 1991), these improved modelling
methods allow for the determination of the shape of dark halos of spiral
galaxies. 

\vspace{0.5cm}

In \S \ref{sec-dark-halo-props} I review some of the properties of dark
halos and determine (in \S \ref{sec-flattened-dark-halos}) how, for a
given rotation curve, the core radius and central density depend on the
flattening of such an isothermal halo (see also Appendix A).  The
disk-halo conspiracy is investigated in section \S
\ref{sec-disk-halo-conspiracy} (and Appendix B) where I derive general
formulas for the three unknowns (the disk's \MoverL ratio and the halo's
core radius and central density) which determine the overall mass
distribution of the galaxy.  In \S \ref{sec-the-works} I list the
assumptions made to determine the z-distribution of the gas from the
galaxian potential.  I compare the local and global approaches in \S
\ref{sec-The-Local-Approach}, where I also present the vertical force
arising from the total mass distribution of the galaxy, $K_{z,tot}$, for
several radii (an analytic solution to the multi-component local
approximation, as proposed by Bahcall (1984), is given in Appendix C). 
In \S \ref{sec-halo-shape-gas-layer-thickness} I calculate how the
thickness of the gas layer depends on the two free parameters (\MoverL
and $q$) and discuss how the thickness of the gas layer can be used to
determine the flattening of the DM-halo.  In the discussion, \S
\ref{sec-discussion}, I will indicate which systems might be suitable
for an analysis as proposed in this paper.

\section{Properties of dark halos}

\label{sec-dark-halo-props}

From the observational fact that rotation curves of spiral galaxies are
``flat'', it has been concluded (e.g.  Bosma 1978; Rubin \etal 1980;
Bahcall \& Casertano 1985; Begeman 1987) that there is an unseen mass
component present in these galaxies.  The standard assumption has been
that this dark mass distribution is spherical and isothermal,
characterized by the core radius, $R_c$, and central density,
$\rho_{h,0}$ (equation [\ref{eq:rho.halo.Rz}], with $q=c/a$=1).  The
luminous matter consists of a stellar disk, for some systems a bulge,
and a gaseous disk.  The \MoverL ratios of the bulge and the disk are
free parameters but are normally taken to be constant with radius.  An
upper limit to the \MoverL ratio of the disk, and hence a lower limit to
the dark halo mass, can be obtained by assuming that the observed peak
rotation velocity is due to the stellar disk only.  A more common
approach is to scale the mass-to-light ratios down in such a way as to
avoid halos with hollow cores.  This is commonly known as the
``maximum-disk'' hypothesis (van Albada \& Sancisi, 1986).  Other means
of constraining the \MoverL ratio exist : van der Kruit (1981)
calculates \MoverL from the thickness of the gas layer, Efstathiou \etal
(1982) invoke disk stability arguments, Athanassoula \etal (1987, ABP87)
apply spiral instability criteria, and Bottema (1993) uses stellar
velocity dispersion measurements.  The mass-to-light ratios found by
these authors range from 50 to 100 \% of the maximum-disk value. 

The values for the DM-halo core radius and the \MoverL ratio of the
stellar disk are highly correlated : low mass-to-light ratios require
small core radii, and large \MoverL ratios correspond to large values
for the DM-halo core radius.  In fact, many different disk-halo
decompositions produce acceptable fits to a given observed rotation
curve (van Albada {\it et al.}, 1985 hereafter ABBS85 ; Persic \&
Salucci, 1988 ; Lake \& Feinswog, 1989 hereafter LF89).  The dark to
luminous mass ratio may decrease with increasing mass (e.g.  ABP87;
Casertano \& van Gorkom 1992; Broeils 1992, Chapter 10).  This could be
inferred from galactic rotation curves : dwarf galaxies have rising
rotation curves (e.g.  Carignan \& Freeman 1988) while the rotation
curves of massive galaxies fall in the outer parts ( Casertano \& van
Gorkom 1991; Broeils 1992).  For those galaxies where the rotation
curves in the outer parts are flat, the luminous and dark matter
``conspire'' : the increase of $V_{halo}^2$ and the decline in
$V_{disk}^2$ are such that their sum remains approximately constant with
galactocentric radius. 

In view of the uncertainties in the disk-halo decomposition, it is clear
that more and a different kind of data are needed for a unique
determination of the halo flattening.  In section \S
\ref{sec-halo-shape-gas-layer-thickness} I show that beyond the optical
disk the thickness of the gas layer is rather sensitive to the
flattening of the halo.  In order to investigate this sensitivity
quantitatively I construct a two parameter galaxy model, where these
parameters are $\gamma$ (= the fraction of the peak rotation curve due
to the stellar disk, see \S \ref{sec-disk-halo-conspiracy} and Appendix
B) and $q$, the flattening of the DM-halo (\S
\ref{sec-flattened-dark-halos} and Appendix A).  The explicit dependence
of the thickness of the gas layer upon these two parameters is discussed
in \S \ref{sec-halo-shape-gas-layer-thickness}.

\subsection{Flattened dark halos}
\label{sec-flattened-dark-halos}

As mentioned above, the equatorial rotation curve (i.e.  the rotation
curve in the plane of the stellar disk) does not constrain the actual
shape of the halo.  In Appendix A I determine a family of flattened
DM-halo models which have the same (to within $1.4$ \%) equatorial but
different polar rotation curves.  Thus, for this family of halo models,
the radial force does not depend significantly on the shape of the
DM-halo, while the vertical force (and hence the thickness of the gas
layer) does.  In Figure \ref{fig:Vrot.q}, I present the rotation curves
for this DM-halo family graphically.  The lower panel shows the rotation
curves for these DM-halo models, while in the top panel I present the
ratio of the flattened to the round DM-halo rotation curve.  Although
the residuals show systematic behavior, the amplitudes ($\leq 1.4 \%$)
are smaller than the routinely obtained observational errors (e.g. 
Begeman 1989, BE89 hereafter; Broeils 1992).  In conclusion : rotation
curves of flattened DM-halos are indistinguishable from their round
equivalents.  This family of flattened DM-halo models is fully specified
by the equations (\ref{eq:Rc.versus.q}) and (\ref{eq:rho0.versus.q})
which relate the core radius and the central density to the flattening
of the DM-halo and is graphically presented in Figure
\ref{fig:Rc.rho0.versus.q}.  For a given rotation curve, flattened
DM-halo models have larger core radii and central densities than their
round equivalents.  Notice that both the central density and the core
radius have an almost linear dependence on the halo flattening for
moderately flattened DM-halos. 

Of course the DM-halo model I have chosen to work with might very well
be different from the true DM halo mass distribution.  However all mass
distributions, with similar rotation curves, share the general feature
that flatter distributions have larger vertical forces.  Therefore, a DM
halo flattening determined using the formalism outlined in this paper
serves as an indicator of the true flattening of the (unknown) dark
matter density distribution. 

\subsection{ On the disk-halo conspiracy}
\label{sec-disk-halo-conspiracy}


As we can measure the light distribution of the stellar disk only, and
have no a priori knowledge of the mass-to-light ratio of stellar disks,
the relative contributions of luminous and dark matter are not known
(ABBS85 and LF89).  However, as I will show in Appendix B, the galaxy
mass model described above is fully determined by {\em one} parameter
regulating the relative importance of stars and dark matter.  Following
Bottema (1993) I choose this parameter, $\gamma$, to be the fraction of
the peak observed rotation curve which is due to the stellar disk
\GAMMA. 

Different choices of $\gamma$ result in quite different values for the
stellar mass-to-light ratio, the halo's central density and core radius. 
As an example, I present the dependence of these three parameters on
$\gamma$ for the galaxy NGC 3198 (rotation curve and optical parameters
were taken from BE89) in Figure \ref{fig:disk.halo.conspiracy},
and algebraically by the equations (\ref{eq:MoverL.versus.gamma}),
(\ref{eq:Rc.versus.gamma}) and (\ref{eq:result.for.rho0}).  In Figure
\ref{fig:disk.halo.conspiracy}, I have also indicated how a 5\%
uncertainty in the slope of the rotation curve affects the results.  To
obtain the core radius and central density of a flattened dark halo, one
uses the equations (\ref{eq:Rc.versus.gamma}) and
(\ref{eq:result.for.rho0}) and multiplies these values by ${\cal C}(q)$
(equation [\ref{eq:Rc.versus.q}]) and ${\cal H}(q)$ (equation
[\ref{eq:rho0.versus.q}]) respectively. 

In agreement with LF89, I find that observed equatorial rotation curves
do not constrain the core radius of the DM-halo (or $\gamma$ in our
terminology) very well.  This is illustrated in Figure
\ref{fig:rotcur.gamma}, where I present the rotation curve of a model
galaxy which resembles the Sc galaxy, NGC 3198\footnote{ The optical
parameters ($L(0)$=207.1 L$_{\odot}$/pc$^2$, $h_R=2.3$ kpc, $z_e=0.23$
kpc), gaseous surface density distribution and rotation curve
(V$_{obs}(2.3h_R)=146.4$, V$_{obs}(10.0h_R)=148.0$,
V$_{gas}(2.3h_R)=13.4$ and V$_{gas}(10.0h_R)=38.0 \;$ km/s) were taken
from BE89 who used Kent's (1987) photometry to calculate the rotation
curve due to the stellar disk.  The stellar disk is truncated at 6$h_R$
(van der Kruit, 1988).  }.  Several acceptable fits, made with different
$\gamma$'s, are shown. 

In conclusion : the parameterization of the disk-halo conspiracy, as
presented in Appendix B, provides a useful way to perform the disk-halo
decomposition to acceptable accuracy, and allows for a straightforward
way to investigate the dependence of the thickness of the gas layer on
the $\gamma$ value (mass-to-light ratio) of the stellar disk.

\section{The method}

\label{sec-the-works}

In this section I derive the vertical distributions of the HI layer from
the potential of the whole galaxy, $\Phi_{galaxy}(R,z)$, and the
equation of hydrostatic equilibrium.  This derivation is subject to
several simplifying assumptions which are listed below :

\begin{itemize} 
\item The system is in steady state,
\item and is azimuthally symmetric.
\item The velocity dispersion tensor of the gas is symmetric and round,
      so that
\item The equation of hydrostatic equilibrium is a good 
      approximation to the vertical Jeans equation.
\item The vertical velocity dispersion of the HI gas is constant 
      (isothermal)  with z-height.
\item Magnetic and cosmic ray pressures are neglected
\end{itemize}

Obviously this is just an approximate description of reality.  Spiral
density waves violate the first two assumptions, while non-thermal
pressure terms (magnetic, cosmic ray heating, ...) are not included in
the equation of hydrostatic equilibrium.  Keeping these assumptions in
mind, I take as the starting point the vertical Jeans equation in
cylindrical coordinates (as usual, $R$, $\theta$, and $z$ denote the
radial, the tangential and the vertical direction respectively) :

\begin{eqnarray}
\frac{d \; \left( \rho(z)  \dispsq{zz}{(R,z)} \right)}{d \; z}  &=&
   \rho(z) \; \Kz{(z)} - 
   \nonumber \\*[3mm]
   && \hspace*{-30mm}\; \ddRoR \left( R \; \rho \dispsq{Rz}{} \right)  -  
   \ddRoT \left( \rho \dispsq{\theta z}{} \right),
\label{eq:Vertical.Jeans}
\end{eqnarray}

\noindent to determine the vertical distribution of the gas ($\rho(z)$). 
Here \Kz{(z)} is the gravitational force (per unit mass) in the
$+z$-direction and $\disp{}{(R,z)}$ the gaseous velocity dispersion. 
Due to the assumption of azimuthal symmetry, the last term in equation
(\ref{eq:Vertical.Jeans}) vanishes.  The $\sigma_{Rz}$-term essentially
measures the tilt of the velocity dispersion ellipsoid, for cylindrical
rotation it is identically zero.  In the most extreme case the velocity
dispersion ellipsoid points towards the galactic center\footnote{I thank
the referee, Phil Maloney, for pointing this out.}. 


At small $z/R$, $\sigma_{Rz}$ is approximately $(\sigma^2_{RR} -
\sigma^2_{zz})z/R$ so that its contribution is expected to be
small\footnote{In the Galaxy, $\sigma_{RR}$ and $\sigma_{zz}$, as
measured by Malhotra's (1994) and Blitz \etal (1984) are almost equal :
7.8 $\pm$ 3 versus 5.7 $\pm$ 1.2 km/s.}.  Thus, the non-diagonal terms
of the velocity dispersion tensor (i.e.  \dispsq{Rz}{} ) will vanish. 
Then, the vertical Jeans equation reduces to the equation of hydrostatic
equilibrium :
 
\begin{eqnarray}
\hspace*{-8mm}
\frac{d \; \left( \rho_{gas}(z)  \dispsq{z,gas}{(R,z)} \right)}{d \; z}  &=&
   \rho_{gas}(z) \; \Kz{(z)}.
   \label{eq:Hydro.equil}
\end{eqnarray}

\noindent With the assumption that the gas is isothermal in the
z-direction it follows that :

\begin{eqnarray}
\hspace*{-8mm}
\dispsq{z,gas}{} \; \frac{d \; \ln{\rho_{gas}(z)} }{d \; z}  &=&
  -\frac{d \; \Phi_{galaxy}(R,z)}{d \; z},
   \label{eq:isothermal.Hydro.equil}
\end{eqnarray}

\noindent so that

\vspace*{-7mm}
\begin{eqnarray}
&& \hspace*{-14mm}
   \rho_{gas}(R,z) = \rho_{gas}(R,0) e^{-\Phi_{galaxy}(R,z)/\dispsq{z,gas}{(R)}}
   \label{eq:rho.gas.z} \\*[3mm]
&& \hspace*{2mm}
   \approx \rho_0 \exp{(-z^2/2W_{gas}^2)} .
   \label{eq:gas.width.approximate}
\end{eqnarray}

\noindent We see that the gaseous density distribution can be calculated
once the gaseous velocity dispersion and the potential of the
galaxy,$\Phi_{galaxy}(R,z)$, are known.  Generally speaking, the
vertical distribution of the gas will not be a ``simple'' function of z. 
Since the observational data does generally not allow for a more
sophisticated analysis than the fitting of Gaussian functions to the
measurements, I choose to fit Gaussians to the model density
distributions (eqn.  [\ref{eq:rho.gas.z}]) as well\footnote{
\label{footnote.width.estimator} Since the {\em shape} of the potential
changes with galactocentric radius, the gaseous density distribution
will also change shape.  Thus, the accuracy of the approximate density
distribution (eqn.  [\ref{eq:gas.width.approximate}]) will also change
with galactocentric radius.  Other indicators for the width of the gas
layer could be used as well.  For example the normalized second moment
of the density distribution, 
$ W_{{\rm MOM2}} = \sqrt{
   \int_{-\infty}^{+\infty}{z^2 \; \rho(z) \; dz } \; / \;
   \int_{-\infty}^{+\infty}{       \rho(z) \; dz }
                      },
$ 
is more sensitive to the high-z parts of the density distribution,
yielding {\em larger} values for the ``width'' of the gas layer.  On the
other hand, the width could be calculated from the Full Width at Half
Maximum ($W_{FWHM} = FWHM/2.35$), a calculation which is more sensitive
to the low-z part of the density distribution, so that {\em smaller}
values for the ``width'' are found.  Significant differences between
these three width measures indicate that the true density distribution
is not Gaussian.  For the toy model discussed below the $W_{FWHM}$ and
$W_{MOM2}$ width estimators can be as much as 20\% larger and smaller
than $W_{gas}$ respectively (inside the optical disk).  Beyond the
optical disk, the gaseous distribution is close to Gaussian.}.  In the
remainder of this paper I will use the dispersion ($W_{gas}$) of this
Gaussian, as defined by eqn [\ref{eq:gas.width.approximate}], as a
measure of the width of the gas layer. 

I incorporate three components in the present study : 1) a stellar disk
with constant scale-height, 2) a flattened isothermal DM-halo, and 3) a
gaseous disk.  The stellar density distribution for the stars must be
close to the one derived from surface brightness measurements by van der
Kruit \& Searle (\\vdKS81a\&b and vdKS82a\&b), namely :

\begin{eqnarray}
&& \hspace*{-1.5cm} \rho_s(R,z)  \nonumber \\*[3mm]
&& \hspace*{-1.5cm}
   = \MoverL \; \rho_s(0,0) \; e^{-R/h_R} \; {\rm sech}^2(z/2z_e) 
   \hspace{5mm} R \leq R_{max} \hspace{2mm}
   \nonumber \\*[3mm]
&& \hspace*{-1.5cm} =
   0.0 \hspace{2mm} \hspace*{4cm}  R > R_{max},
   \label{eq:stars.rho.lum}
\end{eqnarray}


\noindent with $R_{max}=(4.7 \pm 0.7)h_R$, $2z_e \approx
h_R/(4.7\pm1.8)$, and \MoverL the average mass per unit luminosity. 

\onecolumn

Note that Barteldrees \& Dettmar 1993 (BD93) find that the optical disks
are truncated at significantly smaller radii : $R_{max}=(3.0 \pm
0.2)h_R$, $2z_e \approx h_R/(4.0\pm0.3)$.  In external galaxies it is
very hard to observationally distinguish a sech-squared distribution
from an exponential distribution (i.e.  Wainscoat \etal 1989).  For the
Galaxy on the other hand, there is a large body of evidence suggesting
that the vertical distribution can be better represented by an
exponential distribution\footnote{In this paper I do not include any
thick disk component.  Note that these isothermal and exponential
vertical distributions have the same slope at large z-heights.  Thus for
a given surface density and high-$z$ slope, the exponential distribution
has a twice larger value of the midplane density than the isothermal
distribution (see also van der Kruit, 1988).} : $\rho_s(z) \; =
\;\rho_s(0) \; e^{-z/z_e}$ (Gilmore \& Reid 1983; Pritchet 1983; van der
Kruit 1986; Yoshi \etal 1987).  Therefore, I calculate model potentials
for the exponential distribution.  I chose the flattened isothermal
distribution as proposed by SS90 (equation [\ref{eq:rho.halo.Rz}]),
discussed above, to represent the DM-halo mass distribution.

The vertical force of a truncated stellar disk as well as of a flaring
disk is calculated from the potential :

\begin{eqnarray}
\Psi(R,z) \hspace*{-2mm} &=& \hspace*{-2mm} -2 G 
   \int_0^{\infty} r dr \rho(r,0) \int_{-\infty}^{\infty} dz' \rho(r,z')
   \int_0^{\pi} \frac{d\theta}{|\overline{r} -\overline{R}|} \; \; ,
   \label{eq:equation.for.Psi}  \\
K_z(R,z) \hspace*{-2mm} &=& \hspace*{-2mm} -\frac{d}{dz} \Phi(R,z)
   \label{eq:definition.of.Kz} \\
   &=& -2 G
   \int_0^{\infty} \hspace*{-3mm} r dr \rho(r,0) 
   \int_{-\infty}^{\infty} \hspace*{-3mm}  dz' 
   \frac{ \rho(r,z') (z-z') f(R,z,r,z') }
        { \sqrt{(R+r)^2 + (z-z')^2} \left( [R-r]^2 + [z-z']^2 \right) }
   \; \; ,   \label{eq:equation.for.Kz}
\end{eqnarray}

\noindent where $\rho(r,z')$ is the density distribution for which to
calculate the forces.  At the center of the galaxy, $f(0,z,r,z')$ equals
$\pi$, for all other galactocentric radii $f(R,z,r,z')$ is the complete
elliptic integral of the second kind : $f = E(k) = \int_0^{\pi} d\phi
\sqrt{1-k^2 \sin^2{\phi}}$, with $k^2 = 4 r / [(1+r)^2 +(z-z')^2]$. 
With special care for the region around $(R,z)$ this integral can be
solved numerically and takes about 8 seconds per point on a SPARC 2
processor.  I use an exponential for the vertical density
distribution, $\rho(0,z')$, of the stars, while I approximate the
gaseous distribution by a Gaussian.  As neither the vertical
scale-height nor the midplane density, $\rho(r,0)$, are generally
analytic functions, I store their values in tabular form and determine
the interpolated value whenever the integrating routine (QROMB or QROMO,
Press \etal 1990) requires so.  I used the double exponential stellar
disk (Kuijken \& Gilmore, 1989, hereafter KG89, their eqn.  [27]) as a
test case and found that the two methods of calculation agree to within
1 part in 1,000\footnote{This accuracy equals ten times the accuracy
obtained in the integrating routines.}.

\subsubsection{The global approach}
\label{sec-The.global.approach}

As a first step I use the radial part of the potential to determine the
structural parameters of the stellar disk and DM-halo by requiring that
they reproduce the observed equatorial rotation curve.  The radial force
due to the gas is calculated (using the ROTMOD program in GIPSY, van der
Hulst \etal 1992) assuming that it is infinitely thin.  I use the
observed photometry and a constant \MoverL ratio to calculate the
rotation curve due to the stars (also by using ROTMOD).  I then perform
the disk-halo decomposition outlined in Appendix B. 

I use the values for the mass-to-light ratio of the stellar disk and the
core radius and central density of the DM-halo, as found in the
disk-halo decomposition, to calculate the vertical force due to the
stellar disk (equation [\ref{eq:equation.for.Kz}]), and the DM-halo
(SRJF94's equation [6]).  These forces are integrated numerically to
yield the potential, from which the gaseous volume density distribution,
$\rho_{gas}(R,z)$, is calculated (using eqn.  [\ref{eq:rho.gas.z}], an
assumed gaseous velocity dispersion, and an observed surface density
distributions). 

I follow an iterative procedure, where the gaseous density distribution,
calculated from an approximation to the true total potential, is used to
estimate the gaseous contribution to the potential, which must then be
added to the contributions from the stellar disk and DM-halo to yield a
better approximation to the true total potential, from which a better
approximation to true gaseous density distribution can be calculated,
...  etc. 

Thus, the vertical distribution of the gas is not fully consistent with
the Poisson equation

\vspace*{-5mm}
\begin{eqnarray}
4 \pi  G \; \rho_{tot}(R,z) &=&
     -\frac{1}{R  } \; \frac{\partial (R\;F_R)}{\partial R     } \; + \;
      \frac{1}{R^2} \; \frac{\partial^2 \Phi}{\partial \phi^2} \; - \;
      \frac{\partial \Kz{}}{\partial z},
      \label{eq:Poisson}
\end{eqnarray}

\noindent since I take the vertical distribution of the gas layer to be
Gaussian, and because the radial force ($F_R$) due to a thick gas layer
differs from the radial force arising from an infinitely thin disk. 
These inconsistency are expected to be small because of the small
scale-height and low mass of the gaseous disk.  This procedure to
calculate the thickness of the gas layer is what I term the {\em global
approach}.

In order for the global model to be self-consistent, the stellar
velocity dispersion tensor (\dispsq{*;i,j}{(R,z)}, with $ij$ all the
possible coordinate combinations) has to be such that for
the assumed stellar density distribution and the calculated total
potential, the vertical Jeans equation (\ref{eq:Vertical.Jeans}) will be
satisfied.  Thus, every mass model described above makes two
predictions, firstly the radial variation of the width of the gaseous
layer, and secondly the stellar velocity dispersion.  In this paper I
concentrate on the radial variation of the thickness of the gas layer.

\subsection{The Local Approach}
\label{sec-The-Local-Approach}

To date, most authors have determined the distribution of the gas above
the plane from the {\it local} mass densities only.  In this section I
compare this local approach with the global approach described in the
previous section.  To separate variables I will postpone investigating
the effects of the gaseous self-gravity till \S
\ref{sec-Gaseous-self-gravity}.

First I describe how in the local approach the vertical force is
calculated from the local mass densities and gradients in the rotation
curve.  Assuming that, in the region where the gas is found, the
rotation speed does not depend on $z$-height and that the vertical force
does not depend on radius, the Poisson equation (\ref{eq:Poisson})
reduces to the local Poisson equation:

\begin{eqnarray}
   4 \; \pi \; G \; \rho(z) &=&
   - \frac{1}{R  } \; \frac{d (R\;F_R)}{d R }
   - \frac{ d \Kz{}}{dz} .
   \label{eq:Poisson.local}
\end{eqnarray}

\noindent Combining the appropriate equation of hydrostatic equilibrium
(\ref{eq:isothermal.Hydro.equil}) with the equation above
(\ref{eq:Poisson.local}) I find :

\begin{eqnarray}
\dispsq{z,gas}{} \; \frac{d \; \ln{\rho_{gas}(z)} }{d \; z}  &=&
   -4 \; \pi \; G \; \int_0^z dz'
   \left\{
      \rho_{matter}(z') + 
      \left( \frac{-2}{4\;\pi \;G} \right) 
      \left( \frac{V_{rot}}{R} \; \frac{d V_{rot}}{d R}  \right)
    \right\} .
    \nonumber
\end{eqnarray}
\vspace*{-5mm}

\noindent Setting the rotation curve term to $\rho_{rot}$, it follows
that :

\vspace*{-5mm}
\begin{eqnarray}
\dispsq{z,gas}{} \; \frac{d \; \ln{\rho_{gas}(z)} }{d \; z}  &=& \Kz{(z)}
   \hspace{5mm} = \hspace{5mm}
   -4 \; \pi \; G \; \int_0^z dz'
   \left\{ \rho_{matter}(z') + \rho_{rot}(z') \right\} .
   \label{eq:rho.from.local.approx}
\end{eqnarray}

\noindent Negative and positive rotation densities ($\rho_{rot}$)
correspond to rising and falling rotation curves respectively.  The
effect of a rising rotation curve is to increase the thickness of the
gaseous and stellar disks, while in a falling rotation curve a disk will
be thinner relative to the case where $\rho_{rot}=0\;$.  Bottema \etal
(1987) find that the rotation term can contribute significantly in the
rising parts of galactic rotation curves.  For the case of a flat
rotation curve, equation (9) leads to the plane-parallel-sheet case. 
Equation (\ref{eq:rho.from.local.approx}) can be solved either
iteratively or analytically (if more simplifications are introduced) :
the multi-component and Gaussian-equivalent methods described below.

\twocolumn

\subsubsection{The Gaussian equivalent method}
\label{sec-The-Gaussian-equivalent-method}

In order to arrive at the (analytic) Gaussian-equivalent method, we
simplify matters further by assuming that the self-gravity of the gas is
negligible and that the total mass density does not vary significantly
with height above the plane.  I then arrive at :

\vspace*{-8mm}
\begin{eqnarray}
\dispsq{z,gas}{} \; \frac{d \; \ln{\rho_{gas}(z)} }{d \; z}  &=&
   -4 \; \pi \; G \; \rho_{tot}(0) \; z ,
   \nonumber 
\end{eqnarray}
\vspace*{-5mm}

\noindent so that the vertical distribution is Gaussian with dispersion
$W_{gas,G}$ :

\vspace*{-8mm}
\begin{eqnarray}
\rho_{gas}(z) &=& \rho_{gas}(0) \exp{(-z^2/2W_{gas,G}^2)} .
   \nonumber
\end{eqnarray}
\vspace*{-5mm}

\noindent For which the following relation holds :

\vspace*{-8mm}
\begin{eqnarray}
W_{gas,G} &=& \frac{\disp{z,gas}{}}{\sqrt{4 \; \pi \; G \;
   \rho_{tot}(0)}}
   \hspace{1cm} .
   \label{eq:gas.width.Gaussian-equivalent}
\end{eqnarray}
\vspace*{-5mm}

\noindent In the Gaussian-equivalent method one then assumes that the
Gaussian width ($W_{gas,G}$) can be equated with the observed width
\WIDTH, so that the inferred midplane mass density ($\rho_{inf}$) is
given by :

\vspace*{-5mm}
\begin{eqnarray}
\rho_{inf} &=& \frac{1}{4 \pi G} \; 
   \left( \frac{\dispsq{z,gas}{}}{W_{gas}^2} \right) .
   \label{eq:gas.width.MOST.local}
\end{eqnarray}
\vspace*{-5mm}

\noindent Inside the stellar disk, the Gaussian width ($W_{gas,G}$) will
always be smaller then the true width because, for any density
distribution, the volume density decreases with z-height.  The inferred
midplane mass density, as calculated from the true width $W_{gas}$
(which is larger than $W_{gas,G}$) and equation
(\ref{eq:gas.width.MOST.local}), will therefore underestimate the true
midplane mass density.  This effects is illustrated in Figure
\ref{fig:rho.local.approx}, where I plot the ratio of the inferred
midplane mass density ($\rho_{inf}$) and the actual midplane density
($\rho_{tot}$).  For several mass models resembling NGC 3198, this
method recovers $50$ to $110$ \% of the true midplane density. 
Especially for the round halo case ($q=1.0$) we clearly see the effects
of the truncation of the stellar disk at $R=6h_R$.  The truncation of
the stellar disk of NGC 3198 occurs at a relatively large distance
($6h_R$ versus $4.7h_R$ on average, vdKS81a\&b and vdKS82a\&b, and BD93)
so that it's effect may be relatively small for NGC 3198.  On the other
hand, the effects of the truncation of the stellar disk are smaller if
the vertical distribution of the stars is sech$^2$ rather than
exponential (i.e.  less concentrated towards the midplane).  It should
be noted that these results depend somewhat on the width estimator being
used (see footnote \ref{footnote.width.estimator}), for example the
radial variation of $\rho_{inf}/\rho_{tot}$ arrises partly because the
vertical distribution of the gas gradually changes from a distribution
which is strongly centrally peaked (in the inner two scale-lengths) to a
Gaussian (beyond the optical disk). 


Including the rotation density will improve the Gaussian-equivalent
method beyond the truncation of the optical disk (right-hand side
panel).  The radial variation in the density ratio beyond the optical
disk results from small scale structure in the model rotation curve. 
However, this is possible only if the sum of the rotation and DM
densities is larger than zero (see eqn. 
[\ref{eq:gas.width.Gaussian-equivalent}] and Appendix
\ref{sec-Appendix.local.approach}).  For example, this is {\em not} the
case for a maximum-disk model of NGC 3198 in the inner two radial
scale-lengths. 

Although its accuracy varies systematically with radius, the
Gaussian-equivalent method works reasonably well, especially beyond the
stellar disk.

\subsubsection{The plane parallel sheet case}
\label{sec-The-plane-parallel-sheet-case}

The difference between the plane-parallel-sheet case and the global
approach is most easily illustrated by plotting the vertical forces for
the two cases.  In Figure \ref{fig:Kz.all} $\;$ I plot the \Kz{(z)}
forces, as calculated for a NGC 3198-model, for a selection of radii. 
The dashed line is the vertical force due to a plane parallel sheet of
stars with an exponential vertical distribution and a column density
appropriate for the indicated radius.  The open squares represent the
force generated by a radially truncated exponential disk.  The full line
represents the \Kz{(z)} generated by the combined density distribution
of the stellar disk and the DM-halo.  For the combination of parameters,
maximum-disk and round DM-halo (top-left panel), the
plane-parallel-sheet case overestimates the vertical force in the inner
two scale-lengths of the stellar disk, because the plane parallel sheet
is a bad approximation to the fast radially declining surface density
(there is very little surface area at the central surface density). 
Furthermore, the large negative $\rho_{rot}$ associated with the steeply
rising rotation curve (Figure \ref{fig:rotcur.gamma}) decreases the
effective volume density.  Beyond about $2 \; h_R$ the potential due to
the interior stellar disk becomes more important and will eventually
dominate the potential generated by the local column of stars. 
Additional model calculations (top-right panel : $\gamma=0.9, q=0.1$ ;
bottom-left panel : $\gamma=0.6, q=1.0$ ; bottom-right panel :
$\gamma=0.6, q=0.1$) show that the stellar disk has a dominating
contribution to the vertical force (over the extent of the stellar disk)
if and only if the stellar disk is close to ``maximum'' and if the
DM-halo is close to round (see also \S B.4 and Figure
\ref{fig:rhoSTARS.rhoDM}). 

\subsubsection{Inferences}
\label{sec-Inferences}

From the figures \ref{fig:rho.local.approx}, \ref{fig:Kz.all} and
\ref{fig:Kz.halo} we learn the following :
\newline ---At large galactocentric radii ($R \geq 6 h_R$), \Kz{} is
almost linear for the range of models considered here, resulting in an
almost Gaussian vertical distribution for the gas.  Notwithstanding the
low stellar surface densities at these large distances, the contribution
of the stellar disk to the total vertical force can be substantial :
even at eleven radial scale-lengths the stellar disk contributes about
20 \% of the total vertical force (for the round DM-halo cases).  The
self-gravity of the gas (see Figure \ref{fig:Kz.halo}), which is
discussed in more detail in \S \ref{sec-Gaseous-self-gravity}, can be
important (up to 50\% of the total force) in this region as well. 
\newline ---In the transition region ($4 h_R \leq R \leq 6 h_R$) the
contribution of the stellar disk, the gaseous disks, and the dark halo
to the total vertical force can be of comparable magnitude (for the mass
model with maximum disk, round halo, and typical gaseous surface
densities.  See Figure \ref{fig:Kz.halo}).  Stellar disks are truncated
(vdKS81a\&b; vdKS82a\&b; BD93) causing a sharp discontinuity in the
stellar densities.  However the potential as generated by the total
stellar mass distribution will change much less abruptly.  Thus the
local approaches will strongly overpredict the thickness of the gas layer
in this region (see Figure \ref{fig:rho.local.approx}).  Moving away
from this corner ($\gamma=0.9, q=1.0$) in parameter space increases the
halo's contribution to the total potential significantly. 
\newline ---Inside the optical disk, the potential is dominated by the
stellar disk only in case the disk is close to ``maximum'' and the halo
is almost round.  For the other extreme case (minimum-disk+flat-halo),
the potential is completely dominated by the DM-halo.  In the two
remaining corners of parameter space, stellar disk and DM-halo
contribute about equally to \Kz{}.  In all of these cases, the density
distribution of a tracer population will be almost Gaussian as long as
the velocity dispersion of the tracer is so low that most of the tracer
resides at z-heights below the ``knee'' in \Kz{} (at two to three
stellar scale-heights).  For a tracer population like the atomic gas
with a velocity dispersion of roughly 7 km/s, the entire extent of the
optical disk falls in this regime (see Figure
\ref{fig:Wgas.beyond.Rmax}).  Isothermal tracers with a velocity
dispersion larger than two or three times \disp{z,gas}{} can move beyond
the knee in the \Kz{} curve so that their density distribution can no
longer be adequately described by a single ``simple'' function (cf.  the
thick disk of the Milky Way [e.g.  Gilmore \& Reid 1983]). 
\newline ---As expected, both the amplitude and the shape of the
vertical force are almost independent of the $\gamma$ value of the
stellar disk if the DM-halo is highly flattened.  In case of a round
DM-halo, different $\gamma$ values for the stellar disk do result in
different shapes for the \Kz{} curves. 

As the amplitude of the knee in \Kz{} depends linearly on central
surface density (quadratically on $V_{rot,max}$, eqn. 
[\ref{eq:V2.stars.only.2p3hR}]), the thickness of the gas layer will be
smaller in high rotation curve galaxies and larger in dwarfs (see also
Appendix \ref{sec-Four-Special-Cases}). 

Two radial regimes are poorly modeled by the Gaussian-equivalent as well
as the plane-parallel-sheet local approaches : the region where the
rotation curve rises so sharply that ($\rho_{rot}+\rho_{DM} < 0)$, and
the region where the stellar disk truncates.  Special treatment of these
regions can be avoided by using the global approach. For those regions
where the rotation curve is almost flat, which are far away from the
truncation of the stellar disk, the Gaussian-equivalent method works
reasonably well.

\subsection{The self-gravity of the gas}
\label{sec-Gaseous-self-gravity}

Rupen (1991) finds in NGC 891 as well as in NGC 4565, that the HI volume
densities become comparable to the stellar volume densities beyond about
2 radial scale-lengths.  For the Milky Way this is true at the solar
circle.  The self-consistent solution for the multi-component method as
found in Appendix \ref{sec-Appendix.The.Edge.of.the.Optical.Disk} (eqn. 
[\ref{eq:solution.of.zRHO}]) indicates that the importance of the
self-gravity of the gas scales with the ratio of gaseous to stellar
kinetic energy.  At the solar circle this ratio is about 0.1-0.2, so
that the self-gravity of the gas is expected to become increasingly
important at larger radii.

The importance of the self-gravity of the gas is illustrated in Figure
\ref{fig:Kz.halo}, where I plot the \Kz{} due to the total
(stars+DM+gas) potential (the full line), and the forces arising from
the global DM, stellar and gaseous density distributions (dotted, short
dashed, and long dashed lines respectively).  The self-gravity of the
gas has been calculated using the procedure outlined in \S
\ref{sec-The.global.approach}.

\subsubsection{The multi-component method}
\label{sec-The-multi-component-method}

In case $\rho_{rot}$ and the self-gravity of the gas are neglected it is
rather straightforward to show (van der Kruit, 1988) that different
isothermal components subject to the same cylindrically symmetric
gravitational field have vertical distributions which are powers of each
other.  Including $\rho_{rot}$, the self-gravity of the gas, and the
dark matter halo leads to the multi-component method, first used by
Bahcall (1984) to determine the amount of disk-like dark matter in the
solar neighborhood.  In Appendix \ref{sec-Appendix.local.approach}, I
present an analytic solution (eqn.  [\ref{eq:solution.of.zRHO}], which
needs to be solved iteratively) to equation
(\ref{eq:rho.from.local.approx}).  In the same Appendix I also show that
in the multi-component method the distribution of the gas is a power of
the distribution of the stars as well.  In Figure
\ref{fig:Wgas.Global.versus.Local} I compare the model widths calculated
in the multi-component method (where the vertical distribution of the
stellar disk is fixed to a sech-squared , see Appendix
\ref{sec-Appendix.local.approach}) with those calculated using the
global approach.  The top plot shows the percentage difference between
the global and local approaches.  The structure in these difference
curves mostly arrises due to small scale structure in the observed
rotation curve.  Thus, application of the multi-component method could
be misleading.  For example, if the flaring measurements extend to a
radius at which the local approach predicts a local maximum in the
observable width, too round a DM-halo would be inferred.

I used the multi-component method to compare models of NGC 3918 with a
varying amount of gaseous self-gravity.  The results are presented in
Figure \ref{fig:self-gravity-multi-cmp}.  In the left-hand set of panels
I show the width, \WIDTH, of the gas layer as a function of
galactocentric radius for three different halo flattenings (bottom
panel, $q=1.0$; middle panel, $q=0.5$; upper panel, $q=0.1$).  Four
different curves are presented, the different lines corresponds to
differently scaled gaseous surface densities : 1/100, 1, 5, and 10 times
the measured surface density for the full line, the dotted line, the
short dashed line and the long dashed line respectively.  The models are
calculated\footnote{The effect of the scaling of the gaseous surface
density has been taken into account while performing the disk-halo
decomposition.} for a gaseous velocity dispersion of 7.5 km/s and a
stellar disk with a $\gamma$ value of 0.93 (i.e.  about maximum-disk). 
For the scaled gaseous disks I plot (right-hand set of panels) the ratio
of thickness of the scaled disk and the thickness of the
no-self-gravity-disk.  From this plot we see that not including the
gaseous self-gravity will result in an {\em overestimate} of the width
of the gas layer (by approximately 10\% to 20\%).  For the predicted
gas-layer width to equal the observed widths a more massive DM-halo is
required (eqn.  [\ref{eq:rho0.versus.q}]), so that neglecting the
self-gravity of the gas will result (using eqn. 
[\ref{eq:gas.width.MOST.local}]) in a {\em smaller} value (by
approximately 20\% to 40\%) for the halo flattening, $q$.  Thus, in the
local approach as well as in the global approach (see Figure
\ref{fig:Wgas.beyond.Rmax}), the self-gravity of the gas is found to
have a comparable effect on the inferred DM-halo flattening.

From these plots I also conclude that for galaxies with relatively large
gaseous surface densities the thickness of the gas-layer is not
sensitive to the flattening of the dark matter at all.  Thus, the
self-gravity of the gas may be an unimportant or a significant
contributor to the vertical force (for early and late type systems
respectively).

\subsection{Halo shape and gas layer thickness}
\label{sec-halo-shape-gas-layer-thickness}

In order to investigate the dependence of thickness of the gas layer
upon the flattening of the DM-halo and the \MoverL ratio of the stellar
disk, I determined $\rho_{gas}(R,z)$ (and $W_{gas}(R)$ therefrom) for a
range of galaxy models. 

In Figure \ref{fig:Wgas.beyond.Rmax} \ I plot the width, \\
\WIDTH, of the
gas layer in a $\gamma=0.9$ (i.e.  $\sim$ maximum-disk ) disk-halo
combination for several halo flattenings.  The left-hand panel was
calculated for a double exponential disk.  In the middle panel I
incorporated the truncation of the stellar disk (at $R/h_R=6$), while
the self-gravity of the gas is included in the right-hand panel.  The
effect of the self-gravity of the gas is different for the different
disk-halo models, and is strongest between 4 and 8 radial scale-lengths
: between 20\% and 40\% for the NGC 3198 models.  Beyond several radial
scale-lengths the width of the gas layer depends strongly (roughly
proportional to $\sqrt{q}$) on the flattening of the DM-halo.  Inside
the stellar disk, the width of the gas layer is largely independent of
the DM-halo flattening because the potential there is dominated by the
stellar disk.  However, for models with flatter halos and/or
lower-$\gamma$ stellar disks, the halo can be of importance inside the
stellar disk (see \S \ref{sec-The-multi-component-method}, and Figure
\ref{fig:Kz.all}).  It is important to keep in mind that the
self-gravity of the gas is more important for the rounder DM-halo models
(\S \ref{sec-Gaseous-self-gravity} and compare the Figures
\ref{fig:Kz.halo}-a and \ref{fig:Kz.halo}-b).  So the ``signature'' of
the halo flattening is somewhat reduced if the self-gravity of the gas
is important (see also Figure \ref{fig:self-gravity-multi-cmp}).

Is the thickness of the gaseous layer dependent on the mass-to-light
ratio of the disk? In order to answer that question, I performed the
calculations described above for several $\gamma$'s.  The results (for
$q=1.0$) are presented in Figure \ref{fig:Wgas.and.gamma}.  We see that,
as expected, inside the stellar disk a gas-layer immersed in a more
massive stellar disk will be comparatively thinner.  At large
galactocentric radii the thickness of the gas layer becomes virtually
independent of the \MoverL ratio of the stellar disk.  The same
conclusion can be reached from Figure \ref{fig:rho.local.approx}, where
we see that, beyond the optical disk, the derived midplane mass
densities for different \MoverL-disks are nearly equal. 

\subsection{Effects of incomplete modelling}
\label{sec-Effects-of-incomplete-modelling}

It is instructive to see how uncertainties in gaseous velocity
dispersion and uncertainties in the derivation of the vertical force
affect our estimates of the DM-halo flattening.  Another point of
interest is to investigate what differences may arise when using the
local instead of the global approach. 

To summarize the following sections : the errors introduced by using the
Gaussian-equivalent method (typically a 10-30\% too round halo) tend to
cancel out those introduced by neglecting the self-gravity of the gas
(20-60\% too flat a halo). 

\subsubsection{Uncertainties in the velocity dispersion}

In many practical situations it may be that measurements of the gaseous
velocity dispersion are not available.  This severely limits our ability
to determine the flattening of the DM-halo.  In some face-on galaxies
the azimuthally averaged gaseous velocity dispersion is observed to
decline radially from around 12 km/s in the center to remain constant at
7 km/s beyond $R_{25}$ (van der Kruit \& Shostak 1982; Dickey \etal
1990).  In other systems, the region beyond $R_{25}$, as well as the
inter-arm regions appear to have a constant velocity dispersion of 7
km/s, while on the arms values around 13 km/s are reported, ( Shostak \&
van der Kruit 1984; see Kamphuis 1993 for a review).  In summary, the
measurements to date indicate that the gaseous velocity dispersion
beyond the optical disk has a value of around 7 km/s. 

Overestimating the gaseous velocity dispersion by a factor of
$(1+x)$ causes an overestimate of the midplane mass density by a factor
of $(1+x)^2$.  Beyond the optical disk, too flat a DM-halo distribution
would be inferred (by the same factor $(1+x)^2 \;$). 

\subsubsection{Uncertainties in the vertical force}

Here I estimate the effects of neglecting certain components to the
vertical force (e.g.  the self-gravity of the atomic gas, magnetic
forces, or the gravitational force due to some remote point source).  A
mass-model which uses an underestimate of the true vertical force
requires a larger DM density and thus a smaller values of the halo
flattening to reproduce the observed gas layer widths.  Thus,
underestimating the vertical force by a factor $(1+y)$ results in too
small an estimate for the halo flattening (by a factor $[1+y]$).  The
inverse is true when working with an overestimate of the vertical force. 

Neglecting the self-gravity of the neutral atomic gas, which can
typically contribute 30 \% to the vertical force (\S
\ref{sec-Gaseous-self-gravity} and Figure \ref{fig:Kz.halo}), will
result in a 30 \% too small inferred halo flattening.  Since the known
molecular gas content of galaxies rarely exceeds 10\% of the stellar
mass (which contributes roughly 20\% to \Kz{} at large distances, see \S
\ref{sec-The-Local-Approach} and Figure \ref{fig:Kz.all}), and is
confined to the inner parts of the galaxy like the stars, I expect that
neglecting the molecular gas will result in an underestimate of $q$ by
at most 2\%.

\subsubsection{Using the Gaussian-equivalent method}

Application of the Gaussian-equivalent method to an observed set of gas
layer widths underestimates the inferred midplane density by a radially
varying amount (see \S \ref{sec-The-Local-Approach}, and Figure
\ref{fig:rho.local.approx}).  As a result the inferred halos will be too
round and have core radii that are too large.

\section{Discussion and Conclusions}
\label{sec-discussion}

In the previous sections we have seen that a flattened dark halo
imprints a very specific signature on the flaring of the gas-layer
(Figure \ref{fig:Wgas.beyond.Rmax}).  Measurement of this flaring allows
for the determination of the DM-halo flattening. 

Although the global approach is limited by some of the same assumptions
as the local approach (isothermality of the gas, zero non-thermal
pressure terms, constant \MoverL-ratio of the stellar disk, alignment
between DM-halo and stellar disk, ...), it treats some other aspects
much better (radial gradients in stellar surface density, the effects of
the rising rotation curve, the non-cylindrical symmetry of the
potential, ...).  In particular the region where the rotation curve
rises sharply and the region where the stellar disk truncates are better
analyzed using the global approach.  In those parts where the rotation
curve is `flat' and which are not too close to the truncation of the
stellar disk the Gaussian-equivalent and multi-component methods can be
used to reasonable accuracy.  However, the global approach is less
sensitive to small scale irregularities (Figure
\ref{fig:Wgas.Global.versus.Local}).  To obtain reliable results, any
applied method {\em must} include the self-gravity of the gas. 

Therefore, I propose that the more robust global approach (\S
\ref{sec-The.global.approach}) should be used in programs aimed at
determining the flattening of the dark matter halo.  For a set of global
mass distributions with varying halo flattening one can calculate the
flaring behaviour of the gaseous disk.  The DM-halo model exhibiting the
best agreement between calculated and observed gas-layer widths then
yields the halo flattening.  The accuracy of the determined flattening
can be gauged by varying the structural parameters of the individual
mass components.  Since the velocity dispersion of the gas is a crucial
parameter in the analysis, care must be taken to determine its radial
dependence.  This will be done done in a forthcoming paper where I will
analyze the flaring gas layer of thee almost edge-on galaxy NGC 4244. 

\subsection{Suitable systems}
\label{sec-suitable-systems}

Nearby edge-on spiral galaxies with large (extending beyond the optical
disk) neutral hydrogen envelopes are possible targets for studies of
this kind.  Symmetric systems are preferred since it is more likely that
such systems are in equilibrium.  Furthermore suitable systems should
not exhibit strong warping.  For systems with moderate warps, the
kinematical information can be used to disentangle flaring from warping. 
Unfortunately these kind of systems might be rare. 

In order to investigate the possibility of determining the thickness of
the gas-layer in less inclined systems, I simulated HI observations of
two models similar to NGC 3198 (inclined by 72$^o$ w.r.t.  the line of
sight).  In the thin disk model, I constrained the gas-layer to have a
FWHM of 100 pc everywhere, while in the flaring disk model the gas-layer
has a vertical distribution calculated using the multi-component method
(including the self-gravity of the gas, eqn. 
[\ref{eq:solution.of.zRHO}]).  Both models were calculated with a
gaseous velocity dispersion of 7.5 km/s.  For a given set of
observational parameters (15\as$\;$ beam [$\rightarrow$ 700 pc] and 2.56
km/s velocity resolution) I calculate the intensities originating from
all points in the model galaxy for all velocities.  Taking the
normalized second moment with respect to radial velocity yields the
apparent velocity dispersion for all points in the model galaxy.  In
Figure \ref{fig:second.moment} \ I present the two model velocity
dispersion maps.  The systematic differences, of the order of 5 km/s,
arise due to the fact that a line of sight through a flaring disk
samples a wider range in position angle and galactocentric radius than
the line of sight through a thin disk.  In principle, the gaseous
velocity dispersion can be determined after the layer-thickness-effects
are corrected for.  The thickness of the gas layer can be derived by
comparing individual channel maps\footnote{A channel map is an image (of
a galaxy) taken in a ``narrow band filter'' : 2.56 km/s wide in this
case.}.  Contour maps of a particular channel are presented in Figure
\ref{fig:channel.map} : in the top panel the thin disk map is shown, the
middle panel shows the thick disk equivalent while the difference
between the two is contoured in the lower panel.  Due to its larger
vertical scale-height, the projected width of the thick-disk channel map
should be larger than that of the thin-disk, while the average surface
brightness will be reduced.  This signature of the thick disk is easily
recognized in the displayed contour maps.  Furthermore, contrary to
edge-on galaxies, the flaring can be measured in many different channel
maps so that local irregularities (e.g.  spiral streaming motions,
bubbles, ...) are less likely to hamper the determination of the radial
behaviour of the flaring (cf. Olling \& van Gorkom, 1995). 

Therefore I expect that it must be possible to experimentally determine
both the gaseous velocity dispersion and the thickness of the gas-layer
for moderately inclined systems as well. 

\subsection{A control sample}

It might be possible to test the importance of neglected physics in
galaxies where the rotation curve falls in Keplerian manner.  In such
systems the contribution of the dark matter to the global potential is
presumably small, thereby eliminating the gas layer thickness dependence
on the shape of the DM-halo.  As a result the thickness of the gas layer
should depend on the total visible mass of the galaxy and the gaseous
velocity dispersion only.  Some galaxies exhibit falling rotation
curves, albeit not in Keplerian manner (Casertano \& van Gorkom 1991;
Broeils 1992). 

The measured width of the gas layer can then be compared with model
predictions.  If the gas layer is observed to be thicker than predicted,
one might conclude that magnetic and cosmic ray pressures are important. 
On the other hand, a gas layer thinner than predicted indicates that
either the ionizing extra-galactic radiation field is important (e.g. 
Maloney 1993), or that an unseen disk-like DM component is present
(i.e.  the ``clumpuscules'' proposed by Pfenniger \etal 1994a, and
Pfenniger \& Combes 1994b).

\subsection{Summary}

I have investigated the use of accurate determination of the thickness
of the HI-layer in external galaxies.  In combination with constraints
set by the gaseous velocity dispersion and the rotation curve, I find
that in principle it is possible to determine the shape of DM-halos. 
The gaseous self-gravity contributes significantly to the vertical force
field (up to 40\%) but can be modeled accurately using the global
approach (\S \ref{sec-The.global.approach}) outlined in this paper.  The
thus determined halo flattening is independent of the particular choice
of disk-halo decomposition. 

In practice (not necessarily edge-on) systems with extended HI-disks and
regular velocity fields should be used, provided that they are observed
with high angular and velocity resolution.

\acknowledgments

I would like to thank Jacqueline van Gorkom for the support,
encouragement and stimulating discussions during my years in graduate
school.  Michael Rupen's work initiated this line of research : the many
discussions we had over the years and his comments on an earlier version
of this paper were of great help to me.  Furthermore I would like to
thank HongSheng Zao who reminded me of quite a few analytic tricks
applied in Appendix C.  Penny Sackett helped me a lot by providing me
with the relations for \Kz{} and $F_R$ of flattened DM-halos.  I want to
thank Kevin Prendergast for his suggestion to investigate the global
approach and his careful reading of an earlier version of the
manuscript.  David Schiminowich, as an un-prejudiced reader, provided
useful tips for improvement.  I also thank Phil Maloney, the referee,
for some useful suggestions to improve this paper.  And finally I would
like to thank NRAO and the Kapteyn Astronomical Institute for developing
and maintaining the AIPS and GIPSY software packages respectively. 

This work was supported in part through an NFS grant ( AST-90-23254 to J. 
van Gorkom) to Columbia University. 

\normalsize

\clearpage


\onecolumn

\section*{Appendices}

\appendix

\section{Rotation curves of flattened DM-halos}

As mentioned before, it is not possible to constrain the DM-halo
flattening by measurements of the equatorial rotation curve since both
round and flattened mass distributions can generate the same rotation
curve.  In this Appendix I show that the DM-halo model used by
Sackett \& Sparke (1990) :

\begin{eqnarray}
\rho_h(R,z;q) &=& \rho_{h,0}(q) \; 
  \left( \frac{R_c(q)^2}{R_c(q)^2 + R^2 + (z/q)^2}
  \right) \; , \label{eq:rho.halo.Rz} \EqnNum{rho.halo.Rz}
\end{eqnarray}

\noindent (where $R_c, \; \rho_{h,0}$ and $q \;(=c/a)$ are the DM-halo
core radius, central density and flattening respectively) defines a
one-parameter family of halo models with essentially the same equatorial
rotation curve if core radius and central density depend in a
specific way on this one parameter, $q$ (equations
[\ref{eq:Rc.versus.q}] and [\ref{eq:rho0.versus.q}] respectively). 

SRJF94 give a general relation for both the vertical and the radial
force due to such a dark halo (their equation [6]).  Evaluating this
relation in the midplane, I find that the equatorial rotation curve
arising from a flattened DM-halo is given by :

\vspace*{-8mm}
\begin{eqnarray}
V_{halo}^2(R;\rho_{h,0},R_c,q) &=&
    V_{halo}^2(\infty;\rho_{h,0},R_c,q) \times Q(R;R_c,q) \; ,
    \label{eq:V2.R.q}
\end{eqnarray}
\vspace*{-8mm}

\noindent with

\vspace*{-8mm}
\begin{eqnarray}
&& Q(R;R_c,q) \; = \; \nonumber \\*[3mm]
&&
   \left\{ 1 - 
      \left( \frac{g(q)}{\arctan{g(q)}} \right) 
      \left( \frac{  q^2 R_c^2(q)}{R^2+(1-q^2)R_c^2(q) } \right)
      \arctan{ \left( \frac{R^2+(1-q^2)R_c^2(q)}{q^2 R_c^2(q)}\right) }
   \right\} \; ,
   \label{eq:V2.radvar} 
\end{eqnarray}
\vspace*{-8mm}

\noindent and

\vspace*{-5mm}
\begin{eqnarray}
V_{halo}^2(\infty;\rho_{h,0},R_c,q) &=&
   4 \pi G \; \rho_{h,0}(q) \; R_c^2(q) \; f(q) \; ,
   \label{eq:V2.halo.inf}
\end{eqnarray}
\vspace*{-5mm}

\noindent while $f(q)$ and $g(q)$ are given by :

\vspace*{-5mm}
\begin{eqnarray}
&& f(q) =
   \left( \frac{ q \; \arccos{q}}{\sqrt{1-q^2}} \right)   \hspace{3cm}
   g(q) = 
   \left( \frac{\sqrt{1-q^2}}{q} \right) \; .
   \label{eq:f.q}
\end{eqnarray}

\noindent It is easily verified that equation (\ref{eq:V2.R.q}) indeed
yields the familiar \\ $V_h^2(R) = V_h^2(\infty)
\left\{1-(R_c/R)\arctan{(R/R_c)} \right\}$ \ , the rotation curve due to
a round halo. 

Changing the flattening of the halo should not change the halo rotation
curve significantly, that is to say : $V^2_{halo}(R;q) \approx
V^2_{halo}(R;1)$ for all radii.  Noting that $f(1)=1.0$, I find that at
infinity the following relation holds :

\begin{eqnarray}
\rho_{h,0}(1) \; R_c^2(1) &=& 
   \rho_{h,0}(q) \; R_c^2(q) \times f(q) \; .
   \label{eq:rho.Rc.q.versus.1}
\end{eqnarray}

\noindent Requiring that the round-halo rotation curve equals the
flattened-halo rotation curve at $R=R_c(q)$ leads to the following
relation :

\begin{eqnarray}
\hspace{-0.75cm} 
   \left( 
      1 \; - \; \frac{R_c(1)}{R_c(q)} \; \arctan{\frac{R_c(q)}{R_c(1)}}
   \right)
   &=& 
      \left\{ 1 - 
         \left( \frac{g(q)}{\arctan{g(q)}} \right) 
         \left( \frac{q^2}{2-q^2} \right)  
         \arctan{\left( \frac{2-q^2}{q^2} \right) }
      \right\} \; .
   \label{eq:V2.q.v.V2.1}
\end{eqnarray}

\noindent Noting that the RHS of this equation varies between 1.28 and
1.00 for $q \in [0.1,1.0]$, so that $R_c(q)/R_c(1)$ is constrained to
the interval $[1.00,1.242]$, I expand (to second order) the left-hand
and right-hand sides of this equation with respect to $R_c(q)/R_c(1)$
and $q$ respectively.  The resulting quadratic equation is solved to
obtain the core radius of the DM-halo as a function of flattening.  The
central density is then found from equation
(\ref{eq:rho.Rc.q.versus.1}).  After substituting the solutions for
$R_c(q)$ and $\rho_{h,0}(q)$ in the equation for the rotation curve
(equation [\ref{eq:V2.R.q}]) it is found that the rotation curves of
flattened DM-halos reproduce the rotation curve of a round halo to
within 7 \% (for $q=$0.2).  This error is maximal for $R=0$, changes sign
at $R=R_c(q)$ and slowly drops from -2\% at two core radii to 0\% at
infinity.  Since these deviations are large in the region where rotation
curves are determined, such accuracy is not acceptable for our purpose. 
A little experimentation leads to a solution which is more acceptable :
our best approximation ( errors $\leq$ 1.4\%) to the dependence of halo
core radius and central density on its flattening $q$ is graphically
presented in Figure \ref{fig:Rc.rho0.versus.q} and algebraically below :

\begin{eqnarray}
R_c(q) &=& R_c(1) \times
   \left( 1.209 - 0.332 q + 0.123 q^2 \right)
   \nonumber \\*[3mm]
&=&   R_c(1) \times {\cal C}(q) \; ,
   \label{eq:Rc.versus.q}
\end{eqnarray}
\vspace*{-8mm}

\noindent and 

\vspace*{-5mm}
\begin{eqnarray}
\rho_{h,0}(q) &=& \rho_{h,0}(1) \times
   \left( \frac{1}{f(q) {\cal C}(q)^2}    \right) \;
   \left( 0.954 + 0.0533 q - 0.0073 q^2  \right)
   \nonumber \\*[3mm]
&=& \rho_{h,0}(1) \times {\cal H}(q) 
    \hspace{1cm} \approx \hspace{1cm}
    \left( \frac{\rho_{h,0}(1)}{q} \right) \; ,
   \label{eq:rho0.versus.q}
\end{eqnarray}
\vspace*{-3mm}

\noindent where $\lim_{q \rightarrow 1} R_c(q) = R_c(1)$ and $\lim_{q
\rightarrow 1} \rho_{h,0}(q) = \rho_{h,0}(1)$. Substituting
(\ref{eq:rho0.versus.q}), (\ref{eq:Rc.versus.q}) and (\ref{eq:f.q}) into
equation (\ref{eq:rho.halo.Rz}), I find that at large distances the
local DM-halo densities are roughly proportional to $1/f(q) \approx 1/q$.

%
%
\section{On the Disk-Halo conspiracy }
\label{sec-Appendix.disk.halo.conspiracy}

In order to calculate the potential of the galaxy, the mass
distributions of both the stellar disk and the DM-halo need to be
parameterized in a convenient way.  In this section I describe how the
DM-halo structural parameters are related to those of the stellar disk. 
Given the photometry of the stellar disk and the observed rotation
curve, a disk-halo mass model is fully determined by {\em one} parameter
which regulates the relative importance of stars and dark matter. 
Following Bottema (1993) I choose the parameter, $\gamma$, to be the
fraction of the peak observed rotation curve which is due to the stellar
disk \GAMMA.  With such a parameterization of the disk-halo decomposition
available, it is possible to investigate the effects of the stellar
mass-to-light ratio upon the flaring of the gas layer.  In accordance
with van ABBS85 and LF89, I find that a wide range of stellar
mass-to-light ratios ($\gamma$'s) produce acceptable ``fits'' to
observed rotation curves. 


Since the light distribution of the stellar disk is relatively well
known, the first step in tackling the disk-halo conspiracy is the
description of the rotation curve due to the stellar disk.  When
assuming a constant mass-to-light ratio, the stellar density
distribution can be described, to first order, as radially exponential
and vertically thin.  The amplitude of the rotation curve due to such a
thin exponential stellar disk depends only on its central surface
density and radial scale length.  The {\em shape} is fully determined by
the radial density distribution.  Casertano (1983) has shown that both
the amplitude and shape of stellar rotation curves are sensitive to the
thickness of the stellar disk as well as to the truncation radius,
$R_{max}$ (vdKS81a\&b; vdKS82a\&b; BD93).  Therefore, the disk-halo
conspiracy too will vary with the details of the stellar mass
distribution.  While the amplitude of the inner stellar rotation curve
depends rather strongly on the thickness of the stellar disk, the outer
rotation curve is virtually independent thereof.  I will now investigate
the dependence of the stellar rotation curve on the structural
parameters of the stellar disk ($h_R$, $R_{max}$, and $z_e$). 

I calculated stellar rotation curves, using the program ROTMOD in GIPSY
(van der Hulst \etal 1992), for several truncation radii and disk
thicknesses, and vertical distributions.  As it turned out, the rotation
curves of such stellar disks can be parameterized by a rotation speed at
an inner ($R_{inn}$) and an outer ($R_{out}= \; x \; R_{inn}$) radius :

\begin{eqnarray}
V_{rot,stars}^2(R_{inn}) &=&   f_m \; \pi \; G \; h_R \; L(0) \; \MoverL
  \label{eq:V2.stars.only.2p3hR} \\*[3mm]
V_{rot,stars}(R_{out})   &=&   f_x \; V_{rot,stars}(R_{inn}) \; ,
  \label{eq:V2.stars.only.xpyhR}
\end{eqnarray}

\noindent where $L(0)$ is the central surface brightness, \MoverL the
mass-to-light ratio, and $h_R$ the radial scale length of the stellar
disk.  I take $R_{inn} = 2.3h_R$, because a typical stellar disk
(truncated at $4.5h_R$ and $h_R/z_e=0.1$) has its maximum rotation speed
at this distance.  The constants $f_m$ and $f_x$ depend upon the the
vertical distribution and the radial truncation of the stellar disk. 
For the outer radius I used three different values, $6.0h_R$, $8.0h_R$
and $10.0h_R$.  The results are presented in Table 1 below for a range
in parameters covering the observed range in thickness and truncation
radius ( vdKS81a\&b; vdKS82a\&b; Bottema \etal 1987; BD93).

\noindent The listed values are the average of the rotation curves for a
vertically exponential and a vertically sech$^2$ stellar disk.  With
this choice for the two vertical distributions, sech-squared and
exponential, I bracket the suggested range of suggested distributions
(van der Kruit, 1988).  The uncertainty concerning the true vertical
distribution translates directly into uncertainties,$\Delta f$, in $f_m$
and $f_x$.  Experimentally I find, $\Delta f_m \approx (2\Delta f_6)
\approx (2\Delta f_8) \approx (2\Delta f_{10}) = $ 0.6\%, 1.2\%, 1.7\%
and 2.3\% for $z_e/h_R=$ 0.05, 0.10, 0.15 and 0.20 respectively), so
that $f^{exp} = f + \Delta f$ and $f^{sech2} = f - \Delta f$.  Thus, if
the structural parameters of the stellar disk are well known the errors
on the $f$-values are small.  In situations where the thickness of the
stellar disk are not known (as for all non edge-on systems) one must use the
average values, $\overline{f_m}$ and $\overline{f_x}$, which are listed
in the columns labeled ``$z_e/h_R=averaged$''.  The uncertainties on
these averages are typically 4.5, 2.0, 2.0 and 2.0 \%.  If also no
information is available on the location of the truncation radius, then
$f_m$, $f_6$, $f_8$, and $f_{10}$ should be representative of the range
in allowed values : I suggest $f_m = 0.73 \pm 6\%$, $f_6 = 0.68 \pm
10\%$, $f_8 = 0.57 \pm 10\%$ and $f_{10} = 0.50 \pm 10\%$.

\subsection{The mass-to-light ratio of the stellar disk}

Now I turn to the dark halo contribution to the observed rotation
curve.  As usual, the (square of the) observed rotation curve is
generated by the quadratic sum of the constituting mass components :

\begin{eqnarray}
V_{rot,obs}^2(R) &=& 
   V_{rot,stars}^2(R) \; + \; V_{rot,bulge}^2(R) \; + \; V_{rot,gas}^2(R)
   \; + \; V_{rot,halo}^2(R) \; .
\end{eqnarray}

\noindent Now I define the gas-bulge-corrected observed rotation
speed as the square root of (\ref{eq:Cal.V2}). 

\begin{eqnarray}
{\cal V}_{rot,obs}^2(R)&=& 
   V_{rot,obs}^2(R) - V_{rot,bulge}^2(R) - V_{rot,gas}^2(R) \; ,
   \label{eq:Cal.V2}   \\*[3mm]  
&& \hspace{-3cm}{\rm so \; that \; :} \nonumber \\[3mm]
{\cal V}_{rot,obs}^2(R)&=& 
   V_{rot,stars}^2(R) \; + \; V_{rot,halo}^2(R) \; .
   \label{eq:V2.obs}  \\*[3mm]
&& \hspace{-3cm}{\rm Furthermore \; I \; define \;} \gamma 
   {\rm \; and \; } \beta_x 
   {\rm (=the \; slope \; of \; the \; rotation \; curve) \;  : \; } 
   \nonumber \\*[3mm]
\gamma &=& \frac{ V_{rot,stars}(2.3h_R)}{{\cal V}_{rot,obs}(2.3h_R)}
\hspace*{2cm}
\beta_x \; = \; \frac{{\cal V}_{obs}(x \; h_R)}{{\cal V}_{obs}(2.3h_R)}
   \; . \label{eq:gamma.beta.definition}
\end{eqnarray}

\noindent Given the observed radial scale length and central surface
brightness of the stellar disk, the mass-to-light-ratio is a function of
the stellar contribution to the peak rotation curve ($\gamma$) only and
is given by (from the equations [\ref{eq:V2.stars.only.2p3hR}] and
[\ref{eq:gamma.beta.definition}] ) :

\begin{eqnarray}
\MoverL(\gamma) &=& 
   \left( 
      \frac{{\cal V}_{obs}^2(2.3h_R)}{f_m \; \pi G \; h_R \; L(0)}
   \right) \times \gamma^2
\hspace{2mm} = \hspace{2mm}
   10.25 \left( \frac{{\cal V}_{obs,100}^2(2.3h_R)}{f_{m,0.722}h_{R,1}
      L_{100}(0)} \right) 
   \times \gamma^2 \; ,
   \label{eq:MoverL.versus.gamma}
\end{eqnarray}

\noindent where ${\cal V}_{obs,100}(2.3h_R)$ is the gas-bulge-corrected
observed rotation speed at 2.3 radial scale-lengths in units of 100
km/s, $L_{100}(0)$ is the central surface brightness of the stellar disk
in units of 100 $L_{\odot}/$pc$^2$, $h_{R,1}$ is the radial scale-length
in units of 1 kpc and $f_{m,0.722}$ in units of 0.722 (the $f_m$-value
appropriate for a typical stellar disk with $h_R/z_e = 10.0$ and
$R_{max}/h_R = 4.5$). 

\subsection{The core radius of the DM-halo}

For a given observed rotation curve and a calculated stellar rotation
curve {\em shape}, the core radius is determined uniquely by $\gamma$. 
Applying equation (\ref{eq:V2.obs}) at 2.3 and $x \; h_R$, eliminating
the terms depending on the stellar rotation curve, and using equation
(\ref{eq:V2.R.q}) with $q=1$, I find that the following relations hold
:

\begin{eqnarray}
\left( \frac{ (\beta_x^2-(f_x \; \gamma)^2) } {1-\gamma^2} \right)  &=&
   \left( \frac{V_{rot,halo}^2(x \;h_R)}{V_{rot,halo}^2(2.3h_R)} \right)
   \label{eq:Rc.solve}
   \nonumber \\*[3mm]
&\approx&
    a_{x,0} +
    a_{x,1} \left( \frac{R_c}{h_R} \right)   + 
    a_{x,2} \left( \frac{R_c}{h_R} \right)^2 \; .
   \label{eq:RcOhR.gamma.beta}
\end{eqnarray}

\noindent The approximation (\ref{eq:Rc.solve}) is valid for small
ranges in $R_c / h_R$ only.  I list these ranges, together with the
ranges for the left-hand side of (\ref{eq:Rc.solve}) and the appropriate
$a_x$-values in Table 2.  These piecewise approximations recover the
right-hand side of equation (\ref{eq:Rc.solve}) to within 0.2 \%.  No
effort has been made to match the pieces smoothly.

\noindent For any assumed $\gamma$ and measured $\beta_x$ and $f_x$ one
can calculate the left-hand side of equation (\ref{eq:Rc.solve}) and
find the appropriate $a_x$-values in Table 2 from which
${\cal R}_c=R_c/h_R$ follows :

\begin{eqnarray}
\left( \frac{R_c}{h_R} \right)(\gamma;\beta_x,f_x) &\approx& 
   \frac{\left( -a_{x,1} + \sqrt{(a_{x,2})^2 - 4 a_{x,2} ]
      (a_{x,0} - lhs_x) } \right) }
   {2 a_{x,2}}
   \label{eq:Rc.versus.gamma} \\*[3mm]
&=& {\cal R}_c(\gamma) \; .
   \nonumber
\end{eqnarray}

\subsection{The central density of the DM-halo}

\noindent The DM-halo central density is calculated from (\ref{eq:V2.obs}),
(\ref{eq:gamma.beta.definition}) and (\ref{eq:Rc.versus.gamma}). With 

\begin{eqnarray}
{\cal V}_{obs}^2(2.3h_R) &=& 
   \gamma^2 {\cal V}_{obs}^2(2.3h_R) + V_{halo}^2(2.3h_R) 
   \hspace{0.5cm} = \hspace{0.5cm}
   \left( \frac{1}{1-\gamma^2}\right) V_{halo}^2(2.3h_R) \; ,
   \nonumber
\end{eqnarray}
\vspace*{-8mm}

\noindent it follows that

\vspace*{-8.5mm}
\begin{eqnarray}
\rho_{h,0}(\gamma;\beta_x,f_x) &=&  \left( 0.185 \; \MSpccub \right) \;
   \left( \frac{{\cal V}_{obs,100}(2.3h_R)}{ h_{R,1} }
   \right)^2 \times
   \nonumber \\*[3mm]
&& 
   \left\{ \frac{(1-\gamma^2)}{{\cal R}_c(\gamma)^2} \right\} \;
   \left\{
      1 - \left( \frac{{\cal R}_c(\gamma)}{2.3} \right)
          \arctan{\left( \frac{2.3}{{\cal R}_c(\gamma)} \right)}
   \right\}^{-1} \; ,
   \label{eq:result.for.rho0}
\end{eqnarray}

\noindent where ${\cal R}_c$, the normalized core radius of the
DM-halo, depends strongly upon the choice for $\gamma$, and only weekly
upon the shape of the rotation curve (via $\beta_x$) and the shape of the
stellar mass distribution (via $f_x$). 

As an example I present the dependence of the mass-to-light ratio of the
stellar disk (lower right panel of Figure
\ref{fig:disk.halo.conspiracy}), the central density (upper left panel)
and core radius (lower right panel) of the DM-halo on $\gamma$ for NGC
3198 (photometry and rotation curve taken from BE89).  The
error-bars were calculated assuming the most common observational status
: the truncation radius of the stellar disk is known but the thickness
of the stellar disk is not.  In this case, the accuracy with which one
can determine the mass-to-light ratio of the stellar disk and the
central density and core radius of the DM-halo are typically 3, 0.5 and
4 \% respectively.  If the thickness of the stellar disk is also known
(as would be the case for edge-on galaxies), the errors decrease
substantially, while they blow up if neither the truncation radius nor
thickness of the stellar disk are known.  For galaxies which are about
``maximum-disk'', and have more or less flat rotation curves, Figure
\ref{fig:disk.halo.conspiracy} shows that the core radius of the DM-halo
will be 7 $\pm$ (a factor of two) times the radial scale-length of the
stellar disk, not too different from the maximum-disk-fits performed
on real galaxies (Broeils 1992).  Note that while the core radius of the
dark halo approaches infinity as $\gamma$ nears unity , the dark to
luminous mass (upper right panel) is not singular. 

I tested the procedure outlined above on simulated data, where I added
(in quadrature) a stellar and a DM-halo rotation curve to represent a
simulated total rotation curve.  For falling, flat, as well
as for rising simulated total rotation curves I was able to
accurately recover the input parameters, provided that the slope of the
total rotation curve is determined over as large a range in radius
as possible. 

As already pointed out by ABBS85 and LF89, mass models with a range (a
factor of ten in some cases) in stellar mass-to-light ratios are
consistent with observed rotation curves.  I find, attempting to fit
model rotation curves with ``faulty'' $\gamma$ values (or
\MoverL-ratios) that a wide range of stellar mass-to-light ratios is
possible if small ($\approx$ 2 \%) differences between input and
recovered rotation curves are allowed to occur.  The differences between
input and recovered rotation curves increase significantly inside the
inner point and beyond the outer point which were used for to determine
the slope of the rotation curve.  This is illustrated in Figure
\ref{fig:rotcur.gamma} where I show the resultant fits to the observed
rotation curve of NGC3198 (BE89) for several $\gamma$'s.  I
took the outer radius at 10 radial scale lengths.  Although all $\gamma$
values between 0.93 and 0.53 produce similarly acceptable fits, the
low-$\gamma$-disks proposed by Bottema (1993) ``fit'' somewhat better. 
Since the velocity field of the inner parts of NGC 3198 deviates from
circular rotation (Corradi \etal, 1991), the discrepancies between
observed and model curves (in the inner parts) can not be used as an
indication of goodness of fit. 

Thus the method presented above presents us with an easy way to analyse
many possible disk-halo decompositions.  To obtain the core radius and
central density of a flattened dark halo, one uses the equations
(\ref{eq:Rc.versus.gamma}) and (\ref{eq:result.for.rho0}) and multiplies
these values by ${\cal C}(q)$ (equation [\ref{eq:Rc.versus.q}]) and
${\cal H}(q)$ (equation [\ref{eq:rho0.versus.q}]) respectively. 

\subsection{The Edge of the Optical Disk}
\label{sec-Appendix.The.Edge.of.the.Optical.Disk}

Combining the relations for the \MoverL-ratio, the DM-halo core radius
and central density it follows that :

\begin{eqnarray}
\frac{\rho_{stars}(r)}{\rho_{h}(r)} &=& 27.7
   \left( \frac{h_R/z_e}{10}          \right)
   \left( \frac{0.722}{f_m}           \right)
   \left( \frac{\gamma^2}{1-\gamma^2} \right) \; \times 
   \nonumber \\*[3mm]
&& \hspace*{-2.5cm} e^{-r} \times 
   \left\{
      \left( {\cal R}_c(\gamma)^2 + r^2 \right)
      \left[ 1 - \left( \frac{{\cal R}_c(\gamma)}{2.3} \right)
         \arctan{\left( \frac{2.3}{{\cal R}_c(\gamma)} \right)}
      \right]
   \right\} \; ,
   \label{eq:rho.stars.over.rho.DM}
\end{eqnarray}

\noindent where $r=R/h_R$.  Thus the ratio of stellar to DM midplane
densities is independent of the {\em amplitude} of the observed rotation
curve and depends only weakly on the {\em shape} of the rotation curve
(via $\beta_x$), the truncation radius and thickness of the stellar disk
(via $f_m$ and $f_x$ and $h_R/z_e$).  The dependence of
$\rho_{stars}(r)/\rho_{h}(r)$ upon galactocentric radius is graphically
presented in Figure \ref{fig:rhoSTARS.rhoDM} for several $\gamma$ values
(full line $\gamma$=0.9, dotted line $\gamma=0.8$, short dashed line
$\gamma=0.7$, long dashed line $\gamma=0.6$, dash-dotted line
$\gamma=0.5$).  I term the distance at which the disk densities no
longer dominates\footnote{i.e.  become smaller than a few times the DM
density.} the total midplane densities, $R_{dde}$.  The maximum-disk
like models have $R_{dde}$ values ranging from four to seven,
corresponding to the range where some stellar disks are observed to be
truncated (vdKS81a\&b; vdKS82a\&b).  Furthermore we notice that the
luminous-to-dark matter ratio decreases rapidly with decreasing
$\gamma$ and $q$ values. 

Recently Barteldrees \& Dettmar (1993) have found that some spirals have
edges in their light distribution as close in as two radial
scale-lengths. Only for substantially rising rotation curves (say $\beta
\geq 1.3$) and/or small $\gamma$ values does $R_{dde}$ become as small
as two, suggesting that these systems have rising rotation curves and/or
are DM dominated.

\section{A self-consistent solution for the multi-component method}
\label{sec-Appendix.local.approach}

In this Appendix, from the equation of hydrostatic equilibrium
(\ref{eq:rho.from.local.approx}), I derive a self-consistent analytical
solution for the vertical distribution of the gas for the
multi-component method proposed by Bahcall (1984).  The solution
(equation [\ref{eq:solution.of.zRHO}]) is in the form of an integral
equation for $z(\rho_{gas})$, which may be inverted numerically to yield
$\rho_{gas}(z)$.  This equation can be used, in an iterative manner, to
determine the local dark matter density.  The model galaxies presented
in the Figures \ref{fig:self-gravity-multi-cmp}, \ref{fig:second.moment}
and \ref{fig:channel.map} were calculated using this analytic method. 
As it turns out, an analytic solution can be found if one makes the
approximation that the DM-halo density is constant with height above the
plane in the region where the gas is co-spatial with the DM-halo.  For
DM-halos which are not too flattened this is not a bad approximation for
most galactocentric radii.  For the gas as well as for the stars the
following equation holds :

\begin{eqnarray} 
\dispsq{z,i}{} \; \frac{d \; \ln{ \rho_i(z)} }{d \; z}
&=&
   -4 \pi G  \; \int_{0}^{z} 
      \left\{ \rho_{gas}(z') + \rho_{stars}(z') + (\rho_{halo} + \rho_{rot})
      \right\} \; dz' \; .
   \label{eq:rho.from.local.approx.II}
\end{eqnarray}

\noindent The density and dispersion ($\rho_i$ and \dispsq{z,i}{}) can
represent any isothermal component $i$, stars or gas.  Since the right
hand side of (\ref{eq:rho.from.local.approx.II}) is the same for stars
and gas, it follows that the vertical distributions are powers of each
other.  The density distribution has to be normalized such that the
integral equals the surface density.  So I write :

\begin{eqnarray}
\rho_{gas}(z) &=&
   C_g \; \left( \rho_{stars} \right)^{p_{sg}}
   = \; \rho_{gas}(0) \; 
   \left( 
      \frac{\rho_{stars}(z)}{\rho_{stars}(0)} 
   \right)^{p_{sg}} \; ,
   \label{eq:equation.for.rho.gas}  \\*[3mm]
{\rm with } && \nonumber \\*[3mm]
p_{sg} &=& 
   \frac{\dispsq{stars}{}}{\dispsq{gas}{}}  
   \hspace*{9mm} {\rm and \;} C_g {\rm \; such \; that \;} \hspace*{9mm}
   \Sigma_{gas} = \int_{-\infty}^{+\infty} \rho_{gas}(z') \; dz'  \; .
   \nonumber
\end{eqnarray}

\noindent Similar relations can be defined for $\rho_{stars}$. 
Differentiating (\ref{eq:rho.from.local.approx.II}) with respect to z
yields :

\begin{eqnarray}
A_{gas}\; \frac{d^2 \; \ln{ \rho_{gas}(z)}   }{d \; z^2} &=&
   \rho_{gas}(z) + \rho_{stars}(z) + ( \rho_{halo} + \rho_{rot} )
   \label{eq:second.deri} \; , \\*[3mm]
{\rm with} &&      \nonumber   \\*[3mm]
A_{gas} &=& \; - \frac{\dispsq{z,gas}{}}{4 \pi G} \; .
   \label{eq:equation.for.A}
\end{eqnarray}

\noindent Solving for the gaseous component after defining $\alpha(z) =
\ln{\rho_{gas}(z)}$ I rewrite (\ref{eq:second.deri}) :

\begin{eqnarray}
A_{gas}\; \frac{d^2 \; \alpha(z)   }{d \; z^2} &=&
       e^{\alpha}   + 
   C_s \left( e^{\alpha} \right)^{p_{gs}}  + 
   ( \rho_{halo} + \rho_{rot} ) \; .
   \label{eq:equation.for.A.II}
\end{eqnarray}

\noindent Multiplying (\ref{eq:equation.for.A.II}) by
$2\frac{d\alpha}{dz}$ and using the equality : $ \frac{d}{dz} \; \left(
\frac{d\alpha}{dz} \right)^2 \; = \; 2 \;
\frac{d\alpha}{dz} \; \frac{d^2\alpha}{dz^2} $, this equation reduces to
:

\begin{eqnarray}
A_{gas} \; \frac{d}{dz} \; \left( \frac{d\alpha}{dz} \right)^2 &=& 2 \; 
   \frac{d\alpha}{dz} \; 
   \left\{
          e^{\alpha}   + 
      C_s \left( e^{\alpha} \right)^{p_{gs}}  + 
      (\rho_{halo} + \rho_{rot} )
   \right\} \; .
   \nonumber
\end{eqnarray}

\noindent Integrating the RHS with respect to $\alpha$ yields :

\begin{eqnarray}
A_{gas} \; \frac{d}{dz} \; \left( \frac{d\alpha}{dz} \right)^2 &=& 2 \; 
   \frac{d}{dz} \; 
   \left\{
          e^{\alpha}   + 
      \frac{C_s}{p_{gs}} \left( e^{\alpha} \right)^{p_{gs}}  + 
      (\rho_{halo} + \rho_{rot}) \; \alpha + C
   \right\} \; .
   \nonumber
\end{eqnarray}

\noindent For any physical system, the restoring force vanishes at the
midplane (i.e.  the RHS of [\ref{eq:rho.from.local.approx.II}] equals
zero), so that $\left.  \frac{d\alpha}{dz} \right|_{z=0} = 0$. Thus the
integration constant $C$ is given by :

\begin{eqnarray}
C &=& -\rho_{gas}(0) \;
   \left\{ \;
      \left( 1 + \rho'_{s,g} \; p_{sg}  \right) \; + \;
      (\rho'_{h,g} + \rho'_{r,g} ) \; \ln{\rho_{gas}(0)} \;
   \right\} \; ,
   \label{eq:integration.constant.C}
\end{eqnarray}

\noindent where the primed variables are defined as follows :

\begin{eqnarray}
\rho'_{s,g} &=& \frac{\rho_{stars}(0)}{\rho_{gas}(0)} \; ; \; \hspace{0.2cm}
\rho'_{h,g}  =  \frac{\rho_{halo }(0)}{\rho_{gas}(0)} \; ; \; \hspace{0.2cm}
\rho'_{r,g}  =  \frac{\rho_{rot  }(0)}{\rho_{gas}(0)} \; .
\end{eqnarray}

\noindent Thus :

\begin{eqnarray}
   \sqrt{ \frac{-2}{A_{gas}} } \;
   \left\{ - 
      \left(
         \rho_{gas}   + 
         \frac{C_s}{p_{gs}} \left( \rho_{gas} \right)^{p_{gs}}  + 
         (\rho_{halo} + \rho_{rot}) \; \ln{\rho_{gas}} + C
      \right)
   \right\}^{+0.5} \nonumber
\end{eqnarray}
\begin{eqnarray}
\hspace*{2cm} &=&\frac{d\alpha}{dz} \; .
   \nonumber
\end{eqnarray}

\noindent Rearranging terms and integrating yields :

\begin{eqnarray}
   \hspace*{-1cm}
   \sqrt{ \frac{A_{gas}}{-2} } \;
   \int_{\rho_{gas}(0)}^{\rho_{gas}(z)} \frac{d\rho_{gas}}{\rho_{gas}} \;
   \left\{ -
      \left(
         \rho_{gas}   + 
         \frac{C_s}{p_{gs}} \left( \rho_{gas} \right)^{p_{gs}}  + 
         (\rho_{halo} + \rho_{rot}) \; \ln{\rho_{gas}} + C
      \right)
   \right\}^{-0.5} \nonumber
\end{eqnarray}
\begin{eqnarray}
\hspace*{2cm} &=&  z(\rho_{gas}(z)) \; .
   \nonumber
\end{eqnarray}

\noindent Normalizing the gas distribution, i.e.  defining
$y=\frac{\rho_{gas}(z)}{\rho_{gas}(0)}$, inserting the integration
constant $C$ and using (\ref{eq:equation.for.rho.gas}) and
(\ref{eq:equation.for.A}) I find :

\begin{eqnarray}
z_{gas}(Y) &=&
   \sqrt{ \frac{\dispsq{z,gas}{}}{8 \pi G \; \rho_{gas}(0)} } \; \; \times 
   \nonumber \\*[5mm]
&& \hspace{-2cm}  \int_Y^{1} \frac{dy}{y} 
\sqrt{
\left(
   \frac{1}
   {
   (1 + \rho'_{s,g} \; p_{sg} ) -
   (y + \rho'_{sg} \; p_{sg} \; y^{p_{gs}} ) -
   (\rho'_{h,g} + \rho'_{r,g}) \ln(y)
   }
\right)
}    \; .
   \label{eq:solution.of.zRHO}
\end{eqnarray}

\noindent Notice that the term ($\rho'_{s,g} \; p_{sg}$) equals the
ratio of stellar and gaseous kinetic energy in the midplane.  For
vanishing stellar, halo, and rotational densities (i.e. 
$\rho'_{s,g}=\rho'_{h,g}=\rho'_{r,g}=0.0$ so that the gas is fully
self-gravitating), it can be shown that (\ref{eq:solution.of.zRHO}) then
reduces to the familiar self-gravitating solution, $\rho_{gas}(z)
\propto $ sech$^2(z/z_0)$, indeed. 

Investigating equation (\ref{eq:solution.of.zRHO}), we see that there is
no solution possible when the sum of the rotation and DM-halo density is
negative (i.e.  for a steeply rising rotation curve) because at small
fractional densities, $Y$, the logarithmic term goes to $-\infty$ and
hence dominates, leading to a negative term inside the square root. 
Bottema \etal (1987), applying the local approach,
find that the rotation term can contribute significantly in the steeply
rising parts of galactic rotation curves.  Obviously, since galaxies do
not have holes in those regions where the rotation curve rises, at least
one of the assumptions must break down.  Most likely, deviations from
cylindrical symmetry become important and the equation of hydrostatic
equilibrium is no longer a good approximation to the vertical Jeans
equation. 

An iterative procedure is followed to find the vertical density
distribution of the gas.  Starting from initial guesses for stars and
gas, a solution of (\ref{eq:solution.of.zRHO}) is calculated, which
provides a new (normalized) z-distributions for the first component. 
Note that this equation can be inverted to calculate the vertical
distribution of the stars from the z-distribution of the gas merely by
interchanging the $g$ and $s$ indices.  In order to eliminate numerical
instabilities, I chose the calculated component (i.e.  the LHS of
[\ref{eq:solution.of.zRHO}]) to be the component with the smallest
velocity dispersion.  The normalized distribution, $z_{gas}(Y)$ is then
inverted to yield $\rho_{gas}(z)$ which is then scaled so that its
z-integral yields the appropriate surface density.  The stellar density
distribution is then calculated by raising the gaseous distribution to
the power $p_{gs}$ (i.e.  applying the equivalent of equation
[\ref{eq:equation.for.rho.gas}]) followed by the surface density
normalization.  From the new midplane values the next solution of
(\ref{eq:solution.of.zRHO}) is calculated, ...  until consecutive
changes become small.  For a given stellar velocity dispersion, the
stellar vertical density distribution changes in each iteration step. 
In order to keep the thickness of the stellar disk fixed, as suggested
by observation ( vdKS81a\&b; vdKS82a\&b; BD93), I change the stellar
velocity dispersion after each iteration step accordingly.  This
procedure can be easily expanded to include ``N'' isothermal components. 

The local dark matter density can be determined in an iterative manner
as well.  An upper limit to the local dark matter density can be found
by assuming that all other components are massless.  Using the
Gaussian-equivalent method (equation [\ref{eq:gas.width.MOST.local}]),
the measured width and velocity dispersion of the gas layer the upper
limit of the dark matter density is calculated.  A lower limit to the
local dark matter density is of course zero.  For a values of the dark
matter density halfway between the lower and upper limit, the model
gas-layer thickness is calculated and compared with the observed value. 
If the calculated thickness corresponding to this trial DM-density value
is larger than the observed value, then the lower bound is increased to
the trial value.  This procedure is repeated until convergence is
reached.  Model calculations show that beyond the optical disk, it is
possible to accurately recover the input (into eqn. 
[\ref{eq:solution.of.zRHO}]) DM densities from simulated thickness
measurements.

\section{Four Special Cases}
\label{sec-Four-Special-Cases}

Here I describe the gaseous distribution in the vertical direction for
four interesting limiting cases with very different flaring behavior. 
In the first three of those, the rotation curve is assumed to be
constant, while the fourth case represents gas in a region where the
rotation curve is falling.  In all but the self-gravitating case, the
gas is treated as a massless testparticle component.  The results
are listed below, where all widths, \WIDTH, are expressed in kpc. 

\begin{eqnarray}
&& \hspace*{-3.5cm} {\bf Fully \; self-gravitating \; gas :}
   \nonumber \\*[3mm]
W_{gas}^{self}(R) &\approx& 6.74
     \left( \frac{\dispsq{10}{}}{\Sigma_1(R)} \right)  
     \label{eq:self.gravitating} \\*[5mm]
&& \hspace*{-3.5cm} {\bf Gas \; in \; a \; stellar \; disk : } 
   \nonumber \\*[3mm]
W_{gas}^{stars}(R) &\approx& 0.08
     \sqrt{
        \left(\frac{z_{0,400}}{\mu_{100}(0) \; {\cal M/L}_2 } \right)
     } \times
     \disp{10}{} \; \; \exp{\frac{R}{2h_R} }
     \label{eq:in.stellar.disk} \\*[5mm]
&& \hspace*{-3.5cm} {\bf Gas \; in \; a \; (flattened) \; dark \; halo \; 
   mass \; distribution :}
   \nonumber \\*[3mm]
W_{gas}^{halo}(R) &\approx& 
     \sqrt{ q \; \left( \frac{2.436}{1.436\; +\; q} \right) } \times
     \left( \frac{\disp{10}{}}{V(\infty;1)_{halo,100}} 
     \right) \sqrt{ R_{c,10}^2 + ( R_{10}/{\cal C}(q))^2 }
     \label{eq:in.dark.halo} \\*[5mm]
&& \hspace*{-3.5cm} {\bf Gas \; in \; Keplerian \; falloff \; region :} 
   \nonumber \\*[3mm]
W_{gas}^{rot}(R) &=&  0.484 \; 
   \frac{\disp{10}{}}{\sqrt{M_{tot,11}}} \; R_{10}^{\frac{3}{2}}
   \label{eq:gas.width.rho.ROT}
\end{eqnarray}

\noindent where \disp{10}{} is the gaseous velocity dispersion in units
of 10 km/s, $\Sigma_1(R)$ the gaseous surface density in units of solar
masses per pc$^2$, $z_{0,400}$ the scale height of the stars in units of
400 pc, $\mu_{100}(0)$ the central surface brightness in units
of 100 $L_{\odot}/{\rm pc}^2$, ${\cal M/L}_2$ is the number of solar
masses per solar luminosity, $h_R$ is the radial scale length of
the (exponential) stellar disk, $V(\infty;1)_{halo,100}$ the asymptotic
rotation velocity of the round dark halo in units of 100 km/s,
$R_{c,10}$ the core radius of the dark halo in units of 10 kpc,
$R_{10}$, the galactocentric radius in 10 kpc, ${\cal C}(q)$, is the
core radius correction factor arising from the flatness,$q$, of the DM
halo (is of order unity, see equation [\ref{eq:Rc.versus.q}]), and
$M_{tot,11}$ the total mass of the galaxy in units of $10^{11} \;
M_{\odot}$. 

Notice that the thickness of the gas layer in a potential which is
completely dominated by the rotation density is of the same order of
magnitude as a gas layer in a DM-halo potential only.  Such situations
may arise for galaxies where the rotation curve falls in a close to
Keplerian manner.

If one really wants to use the local approach (e.g.  for quick
estimation purposes) instead of the global approach, one might as well
use the approximation to equation (\ref{eq:solution.of.zRHO}) I found to
be accurate to about 10\% over a wide range of parameters. 
$W_{gas}^{approx}$ can be found from :

\begin{eqnarray}
   \left( \frac{1}  {W_{gas}^{approx}}  \right)^2 &=&
   \left( \frac{w_g}{W_{gas}^{self}}    \right)^2 +
   \left( \frac{1}  {W_{gas}^{no-self}} \right)^2 \; ,
   \label{eq:our.approx}  \\*[3mm]
&& \hspace{-2cm} {\rm where } \nonumber \\*[3mm]
   \left( \frac{1}{W_{gas}^{no-self}}    \right)^2 &=&
   \left( \frac{1}{W_{gas}^{stars}}      \right)^2 +
   \left( \frac{w_{hr}}{W_{gas}^{halo}}  \right)^2 +
   \left( \frac{w_{hr}}{W_{gas}^{rot}}   \right)^2 \; ,
   \label{eq:gas.no.self}
\end{eqnarray}
\begin{eqnarray}
&& \hspace{-2cm} { \rm with } \nonumber \\*[3mm]
w_g &=&
   \frac{ 
   \left( 
      1.15 \; \Sigma_{gas} + 1.55 \; 
         (\Sigma_{stars}+\Sigma_{halo}+\Sigma_{rot} )
   \right) 
   }
   { 
   \left( \Sigma_{gas}+\Sigma_{stars}+\Sigma_{halo}+\Sigma_{rot}
   \right) 
   } \; , \nonumber \\*[3mm]
w_{hr} &=&
   \frac{ 
   \left( 
      1.20 \; \Sigma_{stars} + 1.05 \; (\Sigma_{halo}+\Sigma_{rot} )
   \right) 
   }
   { 
   \left( \Sigma_{stars} + \Sigma_{halo} + \Sigma_{rot}
   \right) 
   } \; ,\nonumber \\*[3mm]
&& \hspace{-2cm} {\rm and } \nonumber \\*[3mm]
\Sigma_{halo} &=& \rho_{halo} \times W_{gas}^{no-self}
    \hspace{0.5cm} {\rm and }  \hspace{0.5cm} 
\Sigma_{rot}  = \rho_{rot}  \times W_{gas}^{no-self} \; ,
\nonumber
\end{eqnarray}

\noindent where the relations for the halo and rotation surface
densities are evaluated (eqn.  [\ref{eq:gas.no.self}]) using
$w_{hr}=1.0$.  The widths, $W_{gas}^{i}$, are the widths as found for
the special cases ( according to the equations
[\ref{eq:self.gravitating}], [\ref{eq:in.stellar.disk}],
[\ref{eq:in.dark.halo}] and [\ref{eq:gas.width.rho.ROT}]).  Inside the
stellar disk, the width of the gas layer is mostly determined by the
gas-inside-stellar-disk term, $W_{gas}^{stars}$.  Beyond, the halo-term,
$W_{gas}^{halo}$, dominates.  The self-gravity of the gas perturbs the
dominant term slightly, but with a different numerical coefficient for
the two regimes because of the different form of the un-perturbed
z-distribution functions (sech$^{2p_{sg}}(z/z_0)$ versus Gaussian
respectively).

I have tested this approximation for several cases.  The first case
corresponds to BE89's mass distribution for NGC 3198.  The second case
corresponds to a $\gamma$=0.6-disk in a flattened ($q=0.1$) dark halo,
thereby minimizing the stellar and maximizing the dark halo
contribution.  The third case consists of BE89's halo and stellar mass
model but with the gas surface density multiplied by a factor of ten. 
In the last test case I also multiply the halo densities by a factor of
ten.  In all these cases equation (\ref{eq:our.approx}) recovers the
exact solution to equation (\ref{eq:solution.of.zRHO}) to within 10 \%.

\setcounter{figure}{0}

\clearpage
\twocolumn

\normalsize

\begin{figure}
\caption{
\label{fig:Vrot.q}
The lower panel shows the rotation curves of a family of DM-halo models
with varying flattening ($q=c/a= 1.0, 0.7, 0.3, 0.1)$, central
density ($\rho_{h,0}(q)$, equation [\protect\ref{eq:rho0.versus.q}]) and core
radius ($R_c(q)$, equation [\protect\ref{eq:Rc.versus.q}]).  The fact that all
curves fall on top of each other illustrates that the equatorial rotation
curve alone is not sufficient to determine the shape of the DM-halo.  In
the top panel I present the ratio of the rotation curve of the
flattened to the round DM-halo for the same $q$'s as in the bottom panel. 
Although the residuals show systematic behavior, the amplitudes are
smaller than the routinely obtained observational uncertainties (i.e. BE89)
}
\end{figure}

\begin{figure}
\caption{
\label{fig:Rc.rho0.versus.q}
In this figure I present the $q$-dependence of the DM-halo central
density (top panel, equation [\protect\ref{eq:rho0.versus.q}]) and core radius
(lower panel, equation [\protect\ref{eq:Rc.versus.q}]).  With these functional
forms for $\rho_{h,0}(q)$ and $R_c(q)$, the resultant equatorial DM-halo
rotation curve is almost independent of $q$ (see Figure I).
}
\end{figure}

\onecolumn

\begin{figure}
\caption{
\label{fig:disk.halo.conspiracy}
Here I show the dependence of the DM-halo parameters on the
mass-to-light ratio of the stellar disk.  As an example I used the
parameters valid for NGC 3198 ($L(0)$=207.1 L$_{\odot}$/pc$^2$,
$h_R=2.3$ kpc, $z_e=0.23$ kpc).  The gaseous surface density
distribution and rotation curve (${\cal V}_{obs}(2.3h_R)=145.8$) were
taken from BE89, who used Kent's (1987) photometry to calculate
the rotation curve due to the stellar disk ($f_m=0.899$ and
$f_{10}=0.578$).  In the lower right panel the \MoverL-ratio is
presented as a function of $\gamma$ \GAMMA.
In the upper left and lower left panel I show the DM-halo central
density, $\rho_{h,0}$, and its core radius, $R_c$ respectively.  The
dark-to-stellar mass ratio is presented in the upper right panel.  To
indicate the sensitivity of the halo parameters on the slope of the
rotation curve, I calculated them for three values : the appropriate
$\beta_8$-value for NGC 3198 (0.985) and the $\pm$ 5\% values. 
}
\end{figure}

\begin{figure}
\caption{ 
\label{fig:rotcur.gamma}
In this figure I present fits to the rotation curve of NGC3198 with
different $\gamma$ values.  In the lower panel I present the observed
rotation curve (open squared), the rotation curve due to the HI-disk
(crosses) and several disk-halo combinations [
full line,
gas+($\gamma$=0.9-stellar-disk)+DMhalo,\MoverL=2.96,$\rho_{h,0}=11.2,R_c=6.26$ ;
dotted line,
gas+($\gamma$=0.7-stellar-disk)+DMhalo,\MoverL=1.79,$\rho_{h,0}=92.6,R_c=2.00$ ;
dashed line,
gas+($\gamma$=0.5-stellar-disk)+DMhalo,\MoverL=0.92,$\rho_{h,0}=684,R_c=0.73$
], where the central densities are expressed in units of m\Msun/pc$^3$
and the core radii in units of kpc.
The upper panel shows the velocity difference between observations and
model.  The large $\gamma=0.9$-value corresponds to the ``maximum-disk''
solution.  A better fit is the $\gamma=0.5$-solution (low \MoverL-disk)
close to the value proposed by Bottema (1993).  Furthermore, since the
velocity field of the inner parts of NGC 3198 deviates from axisymmetric
rotation (Corradi \etal, 1991), the discrepancies between observed and
model curves in the inner regions seem irrelevant.  The only free
parameter in the fitting procedure I used (described in Appendix B) is
$\gamma$ \GAMMA.  The observed rotation curve (and the errors, i.e.  the
difference between both sides, are indicated by the error bars), the
rotation curve due to the gaseous disk, the rotation curve due to the
stellar disk and the optical parameters were taken from BE89. 
}
\end{figure}

\begin{figure}
\caption{
\label{fig:rho.local.approx}
For a model galaxy (consistent with the parameters of NGC 3198 as
observed by BE89) we compare the midplane mass densities inferred from
the thickness of the gas layer ($\rho_{{\rm INF}}$, using the local
approach via equation [\protect\ref{eq:gas.width.MOST.local}], where
\disp{(z,gas)}{}=7.5 km/s) with the true stellar midplane densities
($\rho({\rm STARS)_{INP}}$, left-hand set of panels), the true
(stellar+DM-halo) densities ($\rho({\rm STARS+HALO)_{INP}}$, middle set
of panels), and the true (stellar+DM-halo+rotation) densities
($\rho({\rm STARS+HALO+ROT)_{INP}}$, right-hand set of panels).  The
truncation of the stellar disk at 6$h_R$ is easily recognized for almost
all disk-halo models.  The self-gravity of the gas has not been included
in these calculations.  The bottom two panels were calculated for a
round halo, while the top panels corresponds to a flattened ($q=0.1$)
DM-halo.  The lines correspond to models with different $\gamma$ values
(full line : $\gamma=0.9$, dotted line : $\gamma=0.8$, short dashed line
: $\gamma=0.7$ and the long dashed line : $\gamma=0.6$).  For most
models, the ratio of derived-to-input densities (vertical axis) deviates
significantly and systematically from unity so that the
Gaussian-equivalent method can not be used to reliably determine the
midplane volume densities (see also \S
\protect\ref{sec-The-Local-Approach})
}
\end{figure}

\begin{figure}
\caption{
\label{fig:Kz.all}
In this collage of model galaxies, each plot represents the vertical
force, $K_z(z;\gamma,q)$, as a function of height above the plane ($z$). 
The top plots correspond to a $\gamma$=0.9-disk, the lower two plots to
a $\gamma$=0.6-disk.  The left-hand side plots were calculated for a
round ($q=1.0$) halos; the two right most plots for a flat ($q=0.1$)
halos.  The different panels inside each plot correspond to different
radii (in units of $h_R$).  All panels have been calculated for a galaxy
model consistent with the observed properties of NGC 3198 (BE89).  The
open squares represent the vertical force due to a double exponential
stellar disk.  The dashed line is the vertical force valid for the local
approximation.  The full line is the sum of the vertical force from the
stellar disk and DM-halo.  We notice that the maximum-disk plus round
DM-halo combination is the only region in $\gamma,q$-space where the
stellar disk is dominant over the entire extent of the stellar disk ( $R
\lesssim 5 h_R$ ).  For this combination of $\gamma$ and $q$, the
stellar disk contributes significantly ($\approx$ 20 \% at 11 $h_R$) to
the vertical force, even beyond the truncation of the stellar disk (at
6$h_R$).  Furthermore notice, that the {\em shapes} of the vertical
forces generated by a very flattened DM-halo are very similar to the
{\em shapes} of the vertical force in a round DM-halo, but that the {\em
amplitudes} are about twice larger. 
}
\end{figure}


\begin{figure}
\caption{
\label{fig:Kz.halo}
This figure presents the vertical force in the region beyond the optical
disk.  Again, the galaxy model used, stars+DM-halo+gas, is consistent
with the observed properties of NGC 3198 (BE89).  The full line is the
sum of the vertical force due to the stellar disk, the DM-halo and the
gas layer.  The dotted lines represent the $K_z(z)$ due to the flattened
DM-halo only ($q=1.0$ and $0.3$ for the Figures
\protect\ref{fig:Kz.halo}a (left-hand plot) and
\protect\ref{fig:Kz.halo}b (plot on the right-hand side) respectively). 
The vertical force arising due to the (truncated) stellar disk is
represented by the short dashed line.  The long dashed line represents
the self-gravity of the gas, which' contribution to the total vertical
force can be rather important (30-40\%) beyond the optical disk. 
}
\end{figure}

\twocolumn

\begin{figure}
\caption{
\label{fig:self-gravity-multi-cmp}
In this figure I illustrate the dependence of the thickness of the
gas-layer, \WIDTH, upon it's self-gravity.  I used the multi-component
approximation discussed in Appendix C to calculate NGC 3198-like
mass models with varying amounts of gas.  For the mass models with the
differently scaled gaseous surface densities I performed a separate
disk-halo decomposition, all yielding similarly acceptable fits to
the observed rotation curve.  The left-hand set of panels shows (for
$q=0.1$, $q=0.5$, and $q=1$ from top to bottom) the flaring of the
gas-layer for several gas-surface-density scaling factors.  The scale
factor, $s$, is defined such that $\Sigma_{gas,Model} = s
\Sigma_{gas,observed}$ ;
full         line , $s=0.01$ ;
dotted       line , $s=1$    ;
short dashed line , $s=5$    ;
long  dashed line , $s=10$.
In the right-hand set of panels I plot the thickness ratio of scaled surface
density to the zero surface density case.  From these plots it is clear
that the effects of the self-gravity are moderate, but vary
systematically with radius for the observed gaseous surface densities. 
For the upwardly scaled surface densities I find almost no dependence
of the gas-layer width on the halo flattening.
}
\end{figure}

\begin{figure}
\caption{
\label{fig:Wgas.Global.versus.Local}
In the bottom panel I compare the thickness of the gas layer (\WIDTH) as
calculated in the global approach (the lines with the filled circles)
with the thickness as calculated from the multi-component approach (the
lines).  For both approaches a fixed sech$^2(z/(2z_e))$ stellar
distribution, truncated at $R/h_R$=6, was used.  Three different DM-halo
flattenings are shown (from top to bottom : $q=1.0$, full line; $q=0.3$,
dashed line; $q=0.1$, dotted line).  In the top panel I plot the
percentage difference between the two approaches.  The difference
between the global and multi-component method mainly arrises due to
small scale structure in the observed rotation curve, which has been
used to calculate $\rho_{rot}$. 
}
\end{figure} 

\onecolumn

\begin{figure}
\caption{
\label{fig:Wgas.beyond.Rmax}
For a given rotation curve (BE89), I show how the thickness of
the gas layer (\WIDTH) depends on the flattening of the DM-halo
($q=c/a$).  The different line-types/symbols correspond to different
halo flattenings $q$ ;
$q=1.0$ (full  line), 
$q=0.8$ (dotted line), 
$q=0.6$ (short dashed line), 
$q=0.3$ (long  dashed line)), and
$q=0.1$ (filled circles). The model curves presented were calculated for a
gaseous velocity dispersion of 7.5 km/s.  The left-hand side figure was
calculated using a double exponential stellar disk, which is replaced by
a truncated stellar disk for the middle panel.  For the right-hand side
plot I included the truncation of the stellar disk as well as the
self-gravity of the gas.  For these particular mass models the gaseous
self-gravity influences the model widths more strongly than the
truncation of the stellar disk does. 
}
\end{figure} 

\begin{figure}
\caption{
\label{fig:Wgas.and.gamma}
This figure shows how the thickness of the gas layer (\WIDTH) depends on
the mass-to-light ratio of the stellar disk for several galaxy models
(truncated-stellar-disk+{\it round}-DM-halo) consistent with the
observed properties of NGC 3198 (BE89).  The different line types
correspond to different mass-to-light ratios of the stellar disk ;
(\MoverL=2.98, $\gamma=0.9)$ full line, (\MoverL=2.35, $\gamma=0.8)$
dotted line, (\MoverL=1.80, $\gamma=0.7)$ short dashed line,
(\MoverL=1.32, $\gamma=0.6)$ long dashed line.  Inside the optical disk,
the extreme thinness of the HI disk (in absolute measures) will, quite
likely, limit its usefulness for the determination of the \MoverL-ratio
of the stellar disk to the nearest galaxies.  Beyond the optical disk,
the thickness of the gas layer becomes virtually independent of the
mass-to-light-ratio of the stellar disk, so that a determination of the
DM-halo flattening is independent of \MoverL.  As for Figure
\protect\ref{fig:Wgas.beyond.Rmax} the model curves presented were
calculated for a gaseous velocity dispersion of 7.5 km/s. 
}
\end{figure} 

\begin{figure}
\caption{
\label{fig:second.moment}
This figure illustrates one of the observable effects of a flaring
HI-layer.  The apparent gaseous velocity dispersion (i.e.  the second
moment w.r.t.  recession velocity) of two galaxy models is shown here
(the {\em input} dispersion was 7.5 km/s).  The upper panel was calculated
setting the $FWHM$ of the gas layer to 100 pc everywhere, while the
lower panel corresponds to a model run where the vertical distribution
of the HI layer was calculated in a self-consistent manner (cf.  the
procedure outlined in Appendix C for a model with $q=1$).  A flaring
disk imprints a clearly different pattern on the apparent velocity
dispersion map. Since this is a large-scale
pattern, its detection should be relatively easy.  The model
calculations were convolved to a 15\as x15\as ~beam, whereas the
channels are square and 2.56 km/s wide.  For both models I used the
rotation curve and HI surface density distribution of NGC 3198 (BE89).
}
\end{figure} 

\begin{figure}
\caption{
\label{fig:channel.map}
This figure shows that the effects of a flaring HI layer are also
visible in individual channel maps.  The top panel shows a channel map
($|V_{sys}-V_{chan}|$= 104 km/s, smoothed down to 30\as x30\as) for the
thin hydrogen disk described in Figure \protect\ref{fig:second.moment}. 
In the middle panel I show the same channel but now for the flaring-disk
model.  The ``butterfly-wing'' for the thin hydrogen layer model (upper
panel) is clearly narrower and more peaked than its flaring equivalent
(middle panel).  Because the difference between the two models (lower
panel) is significant over a large range in radii, it is expected that
local irregularities (in the velocity field and/or the surface
density distribution) will not be a limitation in determining the
thickness of the hydrogen layer of inclined galaxies. 
}
\end{figure}

\begin{figure}
\caption{
\label{fig:rhoSTARS.rhoDM}
In this figure I present the radial dependence of the ratio of stellar
to dark matter midplane density for several $\gamma$ values (full line
$\gamma=0.9$, dotted line $\gamma=0.8$, short dashed line $\gamma=0.7$,
long dashed line $\gamma=0.6$, dash-dotted line $\gamma=0.5$).  The
left most figure corresponds to a round DM-halo ($q$=1.0), the right most
to a very flattened DM-halo ($q$=0.1) and the figure in the middle was
calculated for a DM-halo of intermediate flattening ($q$=0.4).  This
ratio is calculated using the disk-halo decomposition for a flat
rotation curve presented in Appendix B
([\protect\ref{eq:rho.stars.over.rho.DM}]).  It can be clearly seen that
in maximum-disk like models the stellar density dominates the
midplane mass density out to about four radial scale-lengths, beyond
which the contribution of the stellar density decreases sharply.  For
the low $\gamma$-models, the stellar density never really dominates. 
}

\end{figure}

\clearpage

\begin{flushleft}

Table 1.

\small

\begin{tabular}{|c|r|r|r|r||r|r|r|r||r|r|r|r|} \hline
$\frac{R_{max}}{h_R}$
        &   \mc{z_e/h_R=0.05}   &   \mc{z_e/h_R=0.10}   &  \mc{z_e/h_R=0.15}    \\ \hline
        & $f_m$ & $f_6$ & $f_8$&$f_{10}$& $f_m$ & $f_6$ & $f_8$&$f_{10}$& $f_m$ & $f_6$ & $f_8$ & $f_{10}$  \\ \hline
2.0     & 0.807 & 0.516 & 0.443 & 0.395 & 0.798 & 0.518 & 0.445 & 0.397 & 0.767 & 0.528 & 0.454 & 0.405 \\
2.5     & 0.841 & 0.553 & 0.473 & 0.421 & 0.809 & 0.563 & 0.482 & 0.429 & 0.778 & 0.574 & 0.492 & 0.438 \\
3.0     & 0.774 & 0.612 & 0.522 & 0.464 & 0.748 & 0.623 & 0.531 & 0.472 & 0.722 & 0.633 & 0.540 & 0.480 \\
3.5     & 0.756 & 0.647 & 0.549 & 0.487 & 0.731 & 0.658 & 0.559 & 0.496 & 0.706 & 0.668 & 0.568 & 0.504 \\
4.0     & 0.750 & 0.670 & 0.567 & 0.502 & 0.724 & 0.682 & 0.576 & 0.511 & 0.700 & 0.693 & 0.586 & 0.519 \\
4.5     & 0.747 & 0.688 & 0.578 & 0.511 & 0.722 & 0.700 & 0.588 & 0.520 & 0.697 & 0.711 & 0.598 & 0.529 \\
5.0     & 0.746 & 0.703 & 0.587 & 0.518 & 0.720 & 0.714 & 0.597 & 0.527 & 0.696 & 0.726 & 0.607 & 0.536 \\
5.5     & 0.746 & 0.718 & 0.592 & 0.522 & 0.720 & 0.729 & 0.603 & 0.532 & 0.695 & 0.740 & 0.613 & 0.541 \\
6.0     & 0.745 & 0.734 & 0.597 & 0.525 & 0.720 & 0.743 & 0.607 & 0.535 & 0.695 & 0.753 & 0.617 & 0.544 \\
6.5     & 0.745 & 0.724 & 0.600 & 0.528 & 0.719 & 0.735 & 0.611 & 0.537 & 0.695 & 0.745 & 0.621 & 0.546 \\
7.0     & 0.745 & 0.722 & 0.603 & 0.529 & 0.719 & 0.733 & 0.614 & 0.538 & 0.695 & 0.743 & 0.624 & 0.548 \\
$\infty$& 0.745 & 0.721 & 0.607 & 0.534 & 0.719 & 0.732 & 0.617 & 0.543 & 0.695 & 0.742 & 0.627 & 0.552 \\ \hline
\end{tabular}

\vspace{1cm}

Table 1. (continued)

\normalsize
\begin{tabular}{|c|r|r|r|r||r|r|r|r|} \hline
$\frac{R_{max}}{h_R}$
        &   \mc{z_e/h_R=0.20}   & \mc{z_e/h_R=averaged} \\ \hline
        & $f_m$ & $f_6$ & $f_8$&$f_{10}$&$\overline{f_m}$&$\overline{f_6}$&$\overline{f_8}$&$\overline{f_{10}}$ \\ \hline
2.0     & 0.748 & 0.534 & 0.460 & 0.410 & 0.780 & 0.524 & 0.451 & 0.402 \\
2.5     & 0.749 & 0.584 & 0.501 & 0.446 & 0.794 & 0.569 & 0.487 & 0.434 \\
3.0     & 0.698 & 0.643 & 0.549 & 0.488 & 0.736 & 0.628 & 0.536 & 0.476 \\
3.5     & 0.683 & 0.679 & 0.577 & 0.513 & 0.719 & 0.663 & 0.563 & 0.450 \\
4.0     & 0.677 & 0.703 & 0.596 & 0.528 & 0.713 & 0.687 & 0.581 & 0.515 \\
4.5	& 0.674 & 0.722 & 0.608 & 0.538 & 0.710 & 0.705 & 0.593 & 0.525 \\
5.0     & 0.673 & 0.737 & 0.617 & 0.545 & 0.709 & 0.720 & 0.602 & 0.532 \\
5.5     & 0.672 & 0.751 & 0.623 & 0.550 & 0.708 & 0.735 & 0.608 & 0.536 \\
6.0     & 0.672 & 0.762 & 0.628 & 0.553 & 0.708 & 0.748 & 0.612 & 0.539 \\
6.5     & 0.672 & 0.755 & 0.631 & 0.555 & 0.708 & 0.740 & 0.616 & 0.542 \\
7.0     & 0.672 & 0.754 & 0.634 & 0.557 & 0.708 & 0.736 & 0.619 & 0.543 \\
$\infty$& 0.672 & 0.752 & 0.637 & 0.561 & 0.708 & 0.737 & 0.622 & 0.548 \\ \hline
\end{tabular}

\clearpage

Table 2.

\normalsize
\begin{tabular}{|r|r|r|r||r|}                             \hline 
lhs-range of (\ref{eq:Rc.solve}) &$a_{6,0}$& $a_{6,1}$    & $a_{6,2}$ & $R_c/h_R$-range   \\ \hline
$\in$  (1.00,1.90 ] &  1.000   & 0.4300   &  0.1140    & $\in$ ( 0  , 1.5] \\
$\in$  (1.90,3.04 ] &  0.7363  & 0.7734   & -0.001442  & $\in$ ( 1.5, 3  ] \\
$\in$  (3.04,4.76 ] &  0.2709  & 1.097    & -0.05814   & $\in$ ( 3  , 6  ] \\
$\in$  (4.76,5.81 ] &  1.400   & 0.7439   & -0.03038   & $\in$ ( 6  ,10  ] \\ \hline
lhs-range of (\ref{eq:Rc.solve}) & $a_{8,0}$ & $a_{8,1}$ & $a_{8,2}$ & $R_c/h_R$-range   \\ \hline
$\in$ ( 1.00, 2.10] &  1.000   & 0.4884   &  0.1645    & $\in$ ( 0  , 1.5] \\
$\in$ ( 2.10, 3.71] &  0.7252  & 0.8293   &  0.05622   & $\in$ ( 1.5, 3  ] \\
$\in$ ( 3.71, 6.76] & -0.3281  & 1.506    & -0.05400   & $\in$ ( 3  , 6  ] \\
$\in$ ( 6.76, 9.16] & -0.01879 & 1.4511   & -0.05341   & $\in$ ( 6  ,10  ] \\ \hline
lhs-range of (\ref{eq:Rc.solve}) & $a_{10,0}$& $a_{10,1}$ & $a_{10,2}$ & $R_c/h_R$-range   \\ \hline
$\in$ ( 1.00, 2.23] &  1.000   & 0.5209   &  0.2007    & $\in$ ( 0  , 1.5] \\
$\in$ ( 2.23, 4.20] &  0.7508  & 0.8190   &  0.1103    & $\in$ ( 1.5, 3  ] \\
$\in$ ( 4.20, 8.47] & -0.6623  & 1.702    & -0.02960   & $\in$ ( 3  , 6  ] \\
$\in$ ( 8.47,12.55] & -1.696   & 2.100    & -0.06758   & $\in$ ( 6  ,10  ] \\ \hline
\end{tabular}

\end{flushleft}

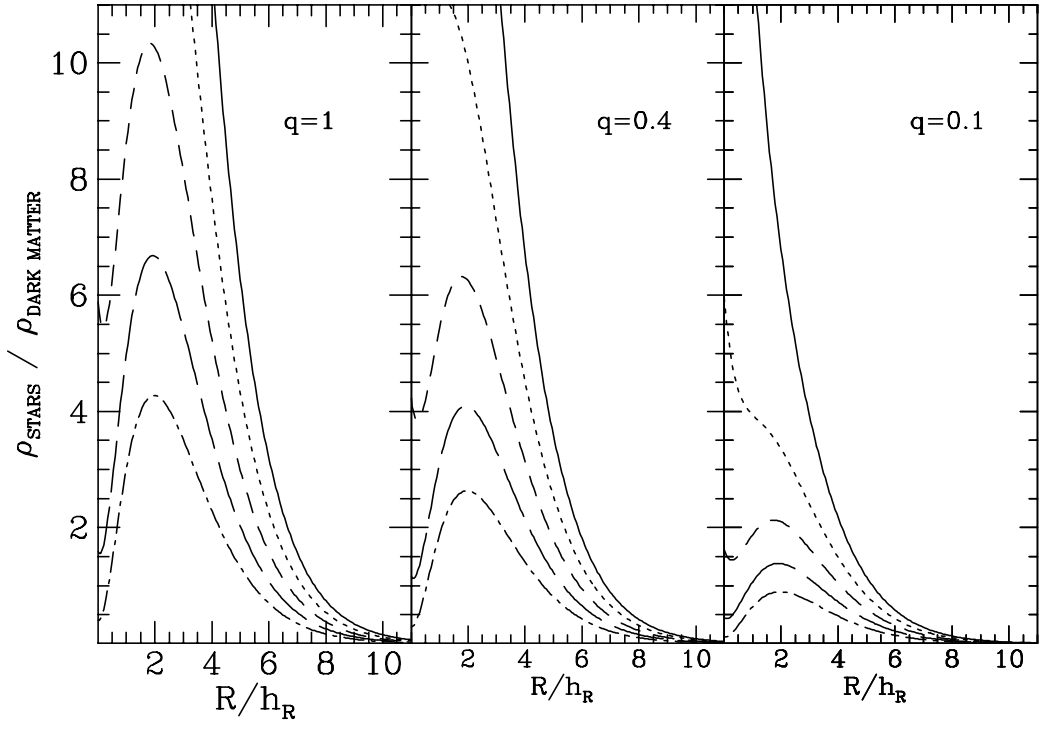
\end{document}